\documentclass[a4paper,11pt]{article}
\pdfoutput=1
\usepackage{jcappub}
\usepackage{bbold}
\usepackage{mathtools}
\usepackage{ulem}

\renewcommand{\thesection}{\arabic{section}}

\def\laq{~\raise 0.4ex\hbox{$<$}\kern -0.8em\lower 0.62ex\hbox{$\sim$}~}
\def\gaq{~\raise 0.4ex\hbox{$>$}\kern -0.7em\lower 0.62ex\hbox{$\sim$}~}

\def\beq{\begin{equation}}
\def\eeq{\end{equation}}
\def\bea{\begin{eqnarray}}
\def\eea{\end{eqnarray}}
\def\bean{\begin{eqnarray*}}
\def\eean{\end{eqnarray*}}

\newcommand{\class}{{\sc class}}

\newcommand{\camb}{{\sc camb}}

\def \ra {\rightarrow}

\def \a {\alpha}

\def\th#1#2{\theta^{(#1)#2}}

\def\laq{~\raise 0.4ex\hbox{$<$}\kern -0.8em\lower 0.62ex\hbox{$\sim$}~}
\def\gaq{~\raise 0.4ex\hbox{$>$}\kern -0.7em\lower 0.62ex\hbox{$\sim$}~}

\def\be{\begin{equation}}
\def\ee{\end{equation}}

\def\beq{\begin{equation}}
\def\eeq{\end{equation}}
\def\bea{\begin{eqnarray}}
\def\eea{\end{eqnarray}}

\def \ra {\rightarrow}

\newcommand{\INT}[1]{\int d^2 \ell_{#1} \ }
\newcommand{\NINT}[1]{\int \frac{d^2 \ell_{#1}}{(2\pi)^2}}
\newcommand{\vl}[1]{\boldsymbol{ \ell}_{#1}}

\newcommand{\T}[2]{\theta^{#1(#2)}}

\newcommand{\Acal}{\mathcal A}

\newcommand{\Mcal}{\mathcal M}

\def\laq{~\raise 0.4ex\hbox{$<$}\kern -0.8em\lower 0.62ex\hbox{$\sim$}~}
\def\gaq{~\raise 0.4ex\hbox{$>$}\kern -0.7em\lower 0.62ex\hbox{$\sim$}~}

\def\beq{\begin{equation}}
\def\eeq{\end{equation}}
\def\bea{\begin{eqnarray}}
\def\eea{\end{eqnarray}}
\def\bean{\begin{eqnarray*}}
\def\eean{\end{eqnarray*}}

\def \ra {\rightarrow}

\def \a {\alpha}

\def\th#1#2{\theta^{#2(#1)}}

\def\th#1#2{\theta^{#2(#1)}}

\title{CMB-lensing beyond the Born approximation}

\author[a,b]{Giovanni~Marozzi}
\author[b,c]{, Giuseppe~Fanizza}
\author[d,e]{, Enea~Di~Dio}
\author[b]{, Ruth~Durrer}

\affiliation[a]{
Centro Brasileiro de Pesquisas F\'{\i}sicas, Rua
  Dr. Xavier Sigaud 150, Urca,  CEP 22290-180, Rio de Janeiro, Brazil
  }
\affiliation[b]{
Universit\'e de Gen\`eve, D\'epartement de Physique Th\'eorique and CAP,
24 quai Ernest-Ansermet, CH-1211 Gen\`eve 4, Switzerland
}
\affiliation[c]{Dipartimento di Fisica, Universit\`a di Bari, 
Via G. Amendola 173, 70126 Bari, Italy, and\\
INFN, Sezione di Bari, Bari, Italy
}
\affiliation[d]{INAF - Osservatorio Astronomico di Trieste, Via
  G. B. Tiepolo 11, I-34143 Trieste, Italy}
  \affiliation[e]{INFN - National Institute for Nuclear Physics,
via Valerio 2, I-34127 Trieste, Italy}

\emailAdd{Marozzi@cbpf.br}
\emailAdd{Giuseppe.Fanizza@ba.infn.it}
\emailAdd{Enea.DiDio@oats.inaf.it}
\emailAdd{Ruth.Durrer@unige.ch}

\abstract{
We investigate the weak lensing corrections to the cosmic microwave background temperature anisotropies considering effects beyond the Born approximation. 
To this aim, we use the small deflection angle approximation, to connect the lensed and unlensed power spectra, via expressions 
for the deflection angles up to third order in the gravitational potential. 
While the small deflection angle approximation has the drawback to be reliable only for multipoles $\ell\lesssim 2500$, it allows us to consistently take into account the non-Gaussian nature of cosmological perturbation theory beyond the linear level. 
The contribution to the lensed temperature power spectrum coming from  the non-Gaussian nature of the deflection angle at higher order is a new effect which has not been taken into account in the literature so far. It turns out to be 
the leading contribution among the post-Born lensing corrections. On the other hand, the effect is smaller than corrections coming from  
 non-linearities in the matter power spectrum, and its imprint on CMB 
lensing is too small to be seen in present experiments.     
}

\begin{document}

\maketitle

\section{Introduction}
\label{Sec1}
\setcounter{equation}{0}
Precise  measurements of the anisotropies and the polarisation of the cosmic microwave background (CMB) have revolutionised cosmology from an order of magnitude science to a precision science. Especially, the latest observations with the Planck satellite~\cite{Adam:2015rua,Planck:2015xua} and also 
from experiments on the ground~\cite{Naess:2014wtr,Crites:2014prc} have determined the cosmological parameters that describe the geometry and the matter content of our Universe and the initial conditions of  perturbations with percent accuracy and better.

To achieve this accuracy it is very important to include the effect of gravitational lensing of CMB photons by foreground structures. Gravitational lensing  has been predicted and calculated to modify the CMB power spectrum by 10\% and more on small angular scales, $\ell\gtrsim 1000$~\cite{1989MNRAS.239..195C,Seljak:1995ve,Zaldarriaga:1998ar,Hu:2000ee,
Lewis:2006fu,RuthBook}. It is therefore imperative to take it into account for precision measurements. In the mean time, lensing of the CMB has not only been detected~\cite{Smith:2007rg,Hirata:2008cb}, but it has been used to reconstruct the lensing potential to the CMB~\cite{Das:2011ak,vanEngelen:2012va,Madhavacheril:2014slf,Story:2014hni} and even a first map of the full sky lensing potential inferred from observations with the Planck satellite has been generated~\cite{Ade:2015zua}. In addition to modifying (damping and widening) the shape of the acoustic peaks in the CMB temperature and polarisation spectra, lensing also rotates E-mode polarisation into B-modes and these lensing B-modes have been detected in the CMB~\cite{Hanson:2013hsb,Ade:2014afa,Keisler:2015hfa,Array:2015xqh,Ade:2015nch}. The possibility to generate high precision maps of the lensing potential to the CMB is intriguing. The lensing potential is a weighted integral of the matter distribution out to the CMB and especially correlating it with different other surveys, for example the infrared background~\cite{Holder:2013hqu,Ade:2013aro,vanEngelen:2014zlh} or galaxy surveys~\cite{Das:2008am,Giannantonio:2015ahz,Kirk:2015dpw,Liu:2016lxy,Baxter:2016ziy}, will give us detailed information about the 
matter distribution in the Universe. For this, however we must ensure that also the theoretical calculation of the CMB deflection angle is sufficiently precise and this is the goal of the present paper.

So far, the effect of lensing in the CMB has been determined using the 'Born approximation', i.e., considering the photons to move along the unperturbed geodesics when computing their deflection angle. This is strictly true only to first order and since lensing is such a significant effect, the question whether higher order terms have to be taken into account  to obtain sufficient accuracy for present and future experiments is justified.  In this paper we study in detail how the CMB temperature power spectrum is affected by lensing including all terms up to third order in the deflection angle. 
For numerical attempts to describe post-Born effects see, for instance, \cite{Hamana:2001vz,Hamana:2004ih,Calabrese:2014gla}.

When going to second order in lensing, also the mean distance out to a fixed redshift is modified. The problem of how the mean luminosity or  area distance of a 2-sphere of constant redshift is affected by aggregated lensing has been extensively studied in recent literature, considering small and intermediate redshift in~\cite{BenDayan:2012ct,BenDayan:2013gc} and going up to the last scattering surface by~\cite{Clarkson:2014pda,Bonvin:2015uha}.
It has been found, however that the effects of this average perturbation on cosmological parameter estimantions are small.

The remaining question, the one we address in this paper, is whether the effects of
'aggregated lensing' on  the CMB temperature fluctuations  power spectrum
are substantial and have to be taken into account. 

Previous estimations of the effect of lensing at higher order on the weak lensing
power spectra found a negligibly small result~\cite{Krause:2009yr}.  However, a recent attempt to evaluate the weak lensing correction of the cosmic microwave background temperature anisotropies at the next to leading order in~\cite{Hagstotz:2014qea} has obtained quite considerable corrections.
The results obtained in~\cite{Hagstotz:2014qea} are however based on the assumption that 
for a stochastic deflection $\a$ the relation $\langle e^{i \a } \rangle=e^{-\langle \a^2 \rangle/2}$ holds. This relation is used when considering first order lensing, to go beyond the small deflection angle approximation.
It allows to re-sum all the 
corrections coming from the first order deflection angle non-perturbatively.
However, it holds only if $\a$ is a Gaussian variable which is 
(nearly\footnote{When taking into account a nonlinear matter distribution in principle also the first order deflection angle is no longer Gaussian. In the present work we focus on the pure post-Born corrections and neglect this non-Gaussianity. However, for the sake of completeness, let us underline that, in principle, the effect 
of the non-Gaussianity associated to the non-linear matter distribution 
can be taken in consideration by expanding $\Phi_W$ as $\Phi_W+\Phi_W^{(2)}+\Phi_W^{(3)}$ in the expressions for the deflection angles. The additional contributions 
are of the same parametric order of the pure post-Born corrections. In our analysis we do not perform these corrections but in some plots we replace the liner power spectrum by a fully non-linear Halofit approximation. In this case, the above Gaussian relation to resum the first order deflection angle can in principle no longer be trusted since then the latter is no longer Gaussian.}) 
 true for the first order deflection angle. But once we go beyond linear order in perturbation theory, the deflection angle no longer obeys  Gaussian statistics and the above identity can not be used.
Actually, as we shall see, using this property corresponds to neglecting terms that are not negligible beyond linear order.
In practice, terms 
of the form $\langle \theta \theta \theta \rangle$, with $\theta$ a deflection angles to any order, cannot be obtained in the above approximation which contains only even powers of  deflection angles. 

For this reason, in the present paper we choose a different approach. We employ the approximation of small deflection angles. With this  
we find here a reliable result which takes into account the Gaussian and non-Gaussian contributions in the lensed CMB anisotropies, going beyond the correction from first order deflection angles.
Starting from the result for the  deflection angle up to third order 
obtained in~\cite{Fanizza:2015swa}, 
we  investigate whether the next to leading order correction coming from foreground structures has to be 
taken in account for a precise calculation of the observed, lensed CMB temperature anisotropy power spectrum.
The results of~\cite{Fanizza:2015swa} are based on the use of the so-called geodesic light-cone gauge~\cite{Gasperini:2011us}. This gauge is especially adapted to the calculation of physical observables and has already previously been applied to this goal, see~\cite{BenDayan:2012wi,Fanizza:2013doa,Marozzi:2014kua,DiDio:2014lka}.

The paper is organized as follow. In Sect.~\ref{Sec2} we present the small deflection angle approximation for CMB lensing beyond linear order, and give the expressions for the deflection angle and the amplification matrix up to 
third perturbative order.
In Sect.~\ref{Sec3} we translate these results into harmonic space, '$\vl{}$--space' and we derive the deflection angle to third order in $\vl{}$--space.
In Sect.~\ref{Sec4} we evaluate the lensed power spectrum of the temperature anisotropies at higher order 
and beyond the Born approximation.
In Sect.~\ref{Sec5} we  introduce the Limber approximation and apply it to some of the results of the previous section.
Our main results are presented in Sect.~\ref{Sec6}. In particular,  we give the Limber approximated expression for the post-Born corrections to the lensed power spectrum of the CMB temperature anisotropies, evaluating numerically the different contribution both considering a linear and a non-linear power spectrum.  
In Sect.~\ref{Sec7} we conclude and add a note regarding a recent paper~\cite{Pratten:2016dsm} that appeared while we 
were finalizing our manuscript.
Some lengthy expressions needed in our calculation in $\vl{}$--space are presented in Appendix~\ref{a:Al}.


\section{Weak Lensing Corrections beyond leading order}
\label{Sec2}
\setcounter{equation}{0}
We want to determine the effect of lensing on the CMB anisotropies beyond the leading order which involves only first order perturbation theory~\cite{Lewis:2006fu,RuthBook}. 
Let us consider a generic scalar field $\Mcal$, that we will identify with the CMB temperature anisotropies, and let us denotes 
 lensed quantities by a tilde ($\tilde{\  }$). 
We can  generalize the result of~\cite{Lewis:2006fu,RuthBook} and write
the following relation between the lensed and unlensed temperature fluctuations $\Mcal$ valid up to fourth order in the deflection angles $\T{a}{i}$ (the superscript $i$ denotes the order).
\bea
\label{eq:expansion}
\tilde\Mcal(x^a)&=&\Mcal\left(x^a+\delta \theta^{a}\right)\simeq\Mcal(x^a)+\sum_{i=1}^4\T{b}{i}\nabla_b\Mcal(x^a)+\frac{1}{2}\sum_{i+j\le 4}\T{b}{i}\T{c}{j}\nabla_{b}\nabla_c\Mcal(x^a)\nonumber\\
&+&\frac{1}{6}\sum_{i+j+k\le 4}\T{b}{i}\T{c}{j}\T{d}{k}\nabla_b\nabla_c\nabla_d\Mcal(x^a) +\frac{1}{24} \T{b}{1}\T{c}{1}\T{d}{1}\T{e}{1}\nabla_b\nabla_c\nabla_d\nabla_e\Mcal(x^a)\,.\nonumber
\\
\label{IniExp}
\eea
This can be written in a more compact form as follows
\bea
\tilde\Mcal(x^a)&\simeq&
\Acal^{(0)}(x^a)+\sum_{i=1}^4\Acal^{(i)}(x^a)
+\sum_{i+j\le 4,\,1\le i\le j}\Acal^{(ij)}(x^a)
+\sum_{i+j+k\le 4,\,1\le i\le j\le k} \Acal^{(ijk)}(x^a)
\nonumber \\
& &+ \Acal^{(1111)}(x^a)
\,,
\label{eq:expansion-compact}
\eea
where
\be
\Acal^{(i_1 i_2....i_n)}(x^a)=\frac{{\rm Perm}(i_1 i_2....i_n)}{n!}\T{b}{i_1}\T{c}{i_2}.....\nabla_{b}\nabla_c.......\Mcal(x^a)
\,,
\ee
where ${\rm Perm}(i_1 i_2....i_n)$ gives the number of permutation of the set $(i_1 i_2....i_n)$,
and \\   $\Acal^{(0)}(x^a)\equiv \Mcal(x^a)$.

We introduce the Weyl potential $\Phi_W$ (in terms of the Bardeen potentials $\Phi$ and $\Psi$) as 
\bea
\Phi_W &=& \frac{1}{2}\left(\Phi+\Psi\right)\,.
\eea
The lensing potential $\psi$ to the last scattering surface
is then given by
\bea
\psi({\bf n},z_s) &=& -\frac{2}{\eta_o-\eta_s} \int_{\eta_s}^{\eta_o} d\eta \frac{\eta-\eta_s}{\eta_o-\eta} \Phi_W(-(\eta_o-\eta){\bf n},\eta)
=-2\int_0^{r_s} dr' \frac{r_s-r'}{r_s r'}\Phi_W(-r'{\bf n},\eta_o-r') \,, \nonumber \\
\label{Lenspot}
\label{Weylpot} 
\eea
with $\bf{n}$ the direction of propagation of photon, $\eta$ conformal time and $r$ the comoving distance, $r=\eta_o-\eta$. The index $_s$ denotes the corresponding quantity at the last scattering surface while $\eta_o$ denotes present time.
The first order deflection angle is simply the gradient of the lensing potential~\cite{Bartelmann:1999yn,RuthBook}. Taking into account also the lensing of the direction $\bf n$ on the path of the photon one
obtains the following expressions for the deflection angles up to third perturbative order~\cite{Fanizza:2015swa}
\bea
\T{a}{1}\!&=&\!
 -2\int_{0}^{r_s}\!dr' \frac{r_s-r'}{r_s\,r'}\nabla^a\Phi_W(r') \,,
\label{TheOrd1}
\\
\T{a}{2}\!&=&\!
-2\int_{0}^{r_s}\!dr'\frac{r_s-r'}{r_s\,r'}
\nabla_b\nabla^a\Phi_W(r')\T{b}{1}(r')\,,
\label{TheOrd2}
\\
\T{a}{3}\!&=&\!
-2\int_{0}^{r_s}\!dr'\frac{r_s-r'}{r_s\,r'}
\!\left[\nabla_b\nabla^a\Phi_W(r')\T{b}{2}(r')\!+\!\frac{1}{2}\nabla_b\nabla_c\nabla^a\Phi_W(r')\T{b}{1}(r')\T{c}{1}(r') \right]\!.
\label{TheOrd3}
\eea
Latin letters $a,b,c,d$ go over the two directions on the sphere.
In Eqs.~\eqref{TheOrd1} to \eqref{TheOrd3} we take into account the terms with the maximal number of  transverse derivatives. We do not expand to higher order the Weyl potential, $\Psi_W$, instead we shall later consider both the linear potential and the fully non-linear one obtained by using Halofit model (see~\cite{Smith:2002dz,Takahashi:2012em}) for the non-linear fluctuations.

Let us also stress that the Taylor expansion in Eq.~(\ref{IniExp}) is valid in the approximation of small deflection angles, i.e. when the deflection angle is much smaller than the angular separations which contribute mainly to $C_\ell$. This is certainly true for $\ell \lesssim 2500$, which corresponds to an angular scale of about $4.5$ arc minutes (see \cite{Seljak:1995ve,RuthBook,Lewis:2006fu}). We shall use the small deflection angle approximation only for the second and third order deflection angles which are much smaller than this value.

To better compare the results that we will obtain in the following, and to understand discrepancy with the ones obtained in \cite{Hagstotz:2014qea}, let us define also the amplification matrix  $\mathcal A^a_b$ following \cite{Fanizza:2015swa}. It is  defined as the derivative of the angular coordinates of the source with respect to the angular direction of the light ray received at the observer position, namely \cite{Bartelmann:1999yn}:
\be 
 \left({ \cal A}^a_b \right)= \left( \frac{\partial \theta_s^a}{\partial \theta_o^b} \right)= 
\left(
\begin{array}{cc}
1 -\kappa - \gamma_1   &  -\gamma_2 - \omega \\
- \gamma_2  + \omega &  1- \kappa+ \gamma_1 
\end{array}
\right) \, .
\label{AmpMat}
\ee
Here $\kappa$ is the convergence, $\gamma_1$ and $\gamma_2$ the two shear components and $\omega$ is the vorticity. As one sees by inspecting Eqs.~(\ref{TheOrd1}) to (\ref{TheOrd3}), the deflection angle $\delta \boldsymbol{\theta }$ can be written as the gradient of a scalar field only to first order. In full generality, introducing a general lensing potential $\varPsi$ and a curl potential $\Omega$,
 we can write the deflection angles $\delta \boldsymbol{\theta}$ (to all order) as
\be
\delta \boldsymbol{\theta }= \boldsymbol{ \nabla} \varPsi + \boldsymbol{\nabla } \times \Omega \, .
\ee
It is then straightforward to show that 
\be
\kappa =- \frac{1}{2}\boldsymbol{ \nabla} \delta \boldsymbol{\theta } = -\frac{1}{2} \Delta \varPsi \qquad, \qquad  \omega =- \frac{1}{2} \Delta \Omega \, 
\ee
and 
\be 
C_\ell^{\varPsi \varPsi}=\frac{4}{\ell^4} C_\ell^{\kappa \kappa}\qquad, \qquad  
C_\ell^{\Omega \Omega}=\frac{4}{\ell^4} C_\ell^{\omega \omega}\,.
\label{SpectraHO}
\ee
Therefore, beyond the Born approximation, the CMB lensed power spectrum will not depend only on a lensing potential, but it has also a curl contribution, see \cite{Hagstotz:2014qea}.
In our approach, by computing directly the deflection angle $\delta \boldsymbol{\theta }$, we do not split the contribution coming from the lensing potential $\varPsi$ and from the curl potential $\Omega$, but we do include them both by construction.
We can then expand in perturbation theory and define the deformation part of the amplification matrix by subtracting the zeroth order contribution, i.e. by introducing the convenient quantity $\Psi^a_b=\delta^a_b -  {\cal A}^a_b$, which is given by \cite{Fanizza:2015swa}
\be
(\Psi^a_b)^{(n)}=-\frac{\partial \theta_s^{a (n)}}{\partial \theta_o^b}, 
~~~~~~~~~~~~ n \geq 1.
\label{GAMordern}
\ee
Using our expressions for the deflection angle we find the following results for the deformation matrix  up to third order (these were first given in 
\cite{Fanizza:2015swa})
\bea 
(\Psi^a_b)^{(1)}&=&2\int_{0}^{r_s} d r' \frac{r_s-r'}{r_s\,r'} \,{\gamma}^{ac}\partial_c
 \partial_b \psi(\eta',r', \theta^a_o) \,,
\\
(\Psi^a_b)^{(2)}&=&2\int_{0}^{r_s} d r' \frac{r_s-r'}{r_s\,r'}  \,{\gamma}^{ac}
\left[\partial_c
 \partial_b \partial_d \psi(r') \theta^{d (1)}-  \partial_c \partial_d \psi(r') \Psi^{d (1)}_b 
 \right] \,,
\\
(\Psi^a_b)^{(3)}&=&2\int_{0}^{r_s} d r' \frac{r_s-r'}{r_s\,r'}  \,{\gamma}^{ac}
\left[\partial_c
 \partial_b \partial_d \psi(r') \theta^{d (2)}+\frac{1}{2} \partial_c \partial_b \partial_d \partial_e \psi(r') \theta^{d (1)}
  \theta^{e (1)} \right.\nonumber \\ 
 & & \left.
 -\partial_c \partial_d \partial_e \psi(r') \theta^{e (1)}  \Psi^{d (1)}_b
   -  \partial_c \partial_d \psi(r') \Psi^{d (2)}_b
 \right] \,,
\eea
where  ${\gamma}_{{ac}}$ denotes the standard metric on the two-sphere and ${\gamma}^{ac}$ is its inverse. The results above are consistent with the ones obtained in~\cite{Krause:2009yr} for the amplification matrix, while they differ from the expressions 
 in~\cite{Hagstotz:2014qea}.
In particular, as already underlined in~\cite{Fanizza:2015swa}, the deformed part of the amplification matrix used 
in~\cite{Hagstotz:2014qea} cannot be obtained as a derivative of a deflection angle order by order, as it should be according to Eq.~(\ref{GAMordern}).
In particular, in~\cite{Hagstotz:2014qea} there seems to be a sign error in the second order expression
$(\Psi^a_b)^{(2)}$, which could be the explanation of an
overestimation of the post-Born corrections to the convergence power spectrum (due to such an error the subtle 
cancellation between the terms $\langle \kappa^{(2)} \kappa^{(2)}\rangle$ and 
$\langle \kappa^{(1)} \kappa^{(3)}\rangle$, which has already been obtained in~\cite{Krause:2009yr} and which also we have recovered  in our framework, might be spoiled).
This could  be the cause of the overestimation of the post-Born correction  of the lensed power spectrum of CMB temperature anisotropies by the authors of \cite{Hagstotz:2014qea}. 
As we  show in the following, the post-Born corrections to 
lensed power spectrum of CMB temperature anisotropies from  
higher order contributions to the convergence (and vorticity) power spectrum are nearly two order of magnitude smaller than the result obtained in \cite{Hagstotz:2014qea}.

\section{Expansion in $\vl{}$-space}
\label{Sec3}

To evaluate the lensing correction to the angular power spectrum $C^{\Mcal}_\ell$ of the CMB temperature anisotropies we consider the flat sky 
limit. In this approximation, see e.g.~\cite{RuthBook}, the combination $(\ell,m)$ is replaced by a two dimensional vector $\vl{}$.
The angular position is then the 2-dimensional Fourier transform of the position in $\vl{}$ space at redshift $z$, for a generic variable $Y(z,{\bf x})$ we have
\be
Y(z,{\bf x})=\frac{1}{2\pi}\INT{}Y(z,\vl{})e^{-i\vl{}\cdot{\bf x}}\,,
\label{Deflspace}
\ee
and 
\bea
\langle Y(z_1,\vl{})\bar{Y}(z_2,\vl{}\,') \rangle&=&\delta\left( \vl{}-\vl{}\,' \right)C_\ell^{Y}(z_1,z_2)\,,
\eea
where an overline denotes complex conjugation.
We denote 
$$ C_\ell^{Y}(z,z)\equiv C_\ell^{Y}(z)\,.$$
To determine  the angular power spectra defined above we follow~\cite{DiDio:2013bqa, DiDio:2014lka} and introduce the (3 dimensional) initial curvature power spectrum by
\be
\langle R_{\rm in} \left( {\bf k} \right) \bar R_{\rm in} \left( {\bf k}' \right) \rangle = \delta_D \left( {\bf k} - {\bf k}' \right) P_R \left(k \right)\,.
\ee
(In both 2- and 3-dimensional Fourier transforms we use the unitary Fourier transform normalisation, hence there are no factors of $2\pi$ in this formula.)

For a given linear perturbation variable $A$ we define its transfer function $T_A(z,k)$ normalized to the initial curvature perturbation by
\be
A \left( z, {\bf k} \right) =T_A(z,k)R_{\rm in}({\bf k}) \,.
\ee
An angular power spectrum will be then given by
\be\label{e:cAB}
C^{AB}_\ell \left( z_1, z_2 \right) = 4 \pi \int \frac{dk}{k} \mathcal{P}_R (k) \Delta^A_\ell (z_1,k)\Delta^B_\ell (z_2,k) = \frac{2}{\pi} \int dk k^2 P_R (k)  \Delta^A_\ell (z_1,k)\Delta^B_\ell (z_2,k) \,,
\ee
where $\mathcal{P}_R (k) = \frac{k^3}{2 \pi^2} P_R (k)$ is the dimensionless primordial power spectrum, and 
$\Delta^A_\ell \left( z, k \right)$ denotes the transfer function in angular and redshift space for the variable $A$. 
For example, by considering $A=B=\Phi_W$ and $A=B=\psi$ we obtain that (setting $C_\ell^{\Psi_W}(z,z')\equiv C_\ell^W(z,z')$)
\bea
\!\!\!\!\!\!\!\!\!\!C_\ell^W(z,z')&=&\frac{1}{2 \pi}\int dk\,k^2\,P_R(k)
\left[ T_{\Psi+\Phi}(k,z)j_\ell\left((k r\right) \right]
\left[ T_{\Psi+\Phi}(k,z')j_\ell\left(k r'\right) \right]\,,
\\
C_\ell^\psi(z,z')&=&\frac{2}{\pi}\int dk\,k^2\,P_R(k)
\left[ \int_{0}^{r}d r_1\frac{r-r_1}{r r_1}
T_{\Psi+\Phi}(k,z_1)j_\ell\left( k r_1 \right) \right] \nonumber \\
& & \times \left[\int_{0}^{r'}d r_2 \frac{ r' -r_2 }{r' r_2}
T_{\Psi+\Phi}(k,z_2)j_\ell\left(  k r_2 \right) \right] \,,
\label{Cpsi}
\eea
where $j_\ell$ denotes a spherical Bessel function of order $\ell$, $r= \eta_o -\eta$ is the comoving distance and analogously for $r', r_1,r_2$. Above and hereafter, we define $z=z(r)$, $z'=z(r')$, etc..

Starting from the definitions in Eq.~(\ref{Deflspace}), we can  write Eq.~(\ref{IniExp}) in $\vl{}$ space in the form
\bea
\tilde\Mcal(z_s,\vl{})&\simeq&
\Acal^{(0)}(\vl{})+\sum_{i=1}^4\Acal^{(i)}(\vl{})
+\sum_{i+j\le 4,\, 1\le i\le j}\Acal^{(ij)}(\vl{})
+\sum_{i+j+k\le 4,\, 1\le i\le j\le k} \Acal^{(ijk)}(\vl{})
\nonumber
\\
& &
+\Acal^{(1111)}(\vl{})\,,
\label{eq:expansion-compact}
\eea
where, on the right hand side, we drop the redshift dependence for simplicity.
In terms of the Fourier transform of the Weyl  potential, the deflections angle up to order three and out to redshift $z_s$ can be written as

\bea
\T{a}{1}({\bf x})
&=&\frac{i}{\pi}
\INT{}\int_{0}^{r_s}dr' \frac{r_s-r'}{r_s\,r'}\,\ell^a
\Phi_W(r',\vl{})e^{-i\vl{}\cdot {\bf x}}
\\
\T{a}{2}({\bf x})
&=&\frac{i}{\pi^2}\INT{1}\INT{2}\int_{0}^{r_s}dr'\frac{r_s-r'}{r_s\,r'}
\left(\,\ell_1^a\ell_{1b}\Phi_W(r',\vl{1})e^{-i\vl{1}\cdot {\bf x}}\right)\nonumber\\
&&\times\int_{0}^{r'}dr'' \frac{r'-r''}{r'\,r''}\,\ell_2^b
\Phi_W(r'',\vl{2})e^{-i\vl{2}\cdot {\bf x}} 
\eea
\bea
\T{a}{3}({\bf x})
&=&\frac{i}{\pi^3}\int_{0}^{r_s}dr'\frac{r_s-r'}{r_s\,r'}\INT{1}\INT{2}\INT{3}\nonumber\\
&&\times\left[ \ell^a_1\ell_{1c}\ell_2^c\ell_{2d}\ell_3^d\Phi_W(r',\vl{1})\int_{0}^{r'}dr''\frac{r'-r''}{r'\,r''}
\Phi_W(r'',\vl{2})\right.\nonumber\\
&&\left.\times
\int_{0}^{r''}dr''' \frac{r''-r'''}{r''\,r'''}\,
\Phi_W(r''',\vl{3})\right. \nonumber\\
&&+\frac{1}{2}\ell^a_1\ell_{1c}\ell_{1d}\ell_2^c\ell_3^d\Phi_W(z',\vl{1})
\int_{0}^{r'}dr'' \frac{r'-r''}{r'\,r''}\,
\Phi_W(r'',\vl{2})\nonumber\\
&&\left.\times\int_{0}^{r'}dr''' \frac{r'-r'''}{r'\,r'''}\,
\Phi_W(r''',\vl{3}) \right]e^{-i\left( \vl{1}+\vl{2}+\vl{3}\right)\cdot {\bf x}}
\eea
Using these results one easily
obtains the $\vl{}$ space expressions for the  terms $\Acal^{(i....)}$. We present all of them in Appendix~\ref{a:Al}.

\section{Lensed angular power spectrum: analytical results}
\label{Sec4}
\setcounter{equation}{0}
Let us now compute the lensed angular power spectrum $\tilde{C}^{\Mcal}_\ell$ using the results of the previous section and of Appendix \ref{a:Al}.
First of all, one can easily  see that
\bea
\langle \tilde{\Mcal}(\vl{})\bar{\tilde{\Mcal}}(\vl{}\,')\rangle =\langle \Acal (\vl{}) \bar{\Acal}(\vl{}\,')\rangle \,,
\eea
where we set
\be
\Acal(\vl{})= \Acal^{(0)}(\vl{})+
\sum_{i=1}^4\Acal^{(i)}(\vl{})
+\sum_{i+j\le 4,\, 1\le i\le j}\Acal^{(ij)}(\vl{})
+\sum_{i+j+k\le 4,\, 1\le i\le j\le k} \Acal^{(ijk)}(\vl{})+\Acal^{(1111)}(\vl{})\,.
\ee
We now introduce  $C_\ell^{(i\ldots ,\, j\ldots)}$ defined as follows
\bea
\delta\left( \vl{}-\vl{}\,' \right) C_\ell^{(ij\ldots,ij\ldots)} &=&\langle \Acal^{(ij\ldots)} (\vl{}) 
\bar{\Acal}^{(ij\ldots)}(\vl{}\,')\rangle\,,
\nonumber \\
\delta\left( \vl{}-\vl{}\,' \right) C_\ell^{(ij\ldots,i'j'\ldots)} &=&
\langle \Acal^{(ij\ldots)} (\vl{}) 
\bar{\Acal}^{(i'j'\ldots)}(\vl{}\,')\rangle+\langle \Acal^{(i'j'\ldots)} (\vl{}) 
\bar{\Acal}^{(ij\ldots)}(\vl{}\,')\rangle\,,
\label{HigherOrderCll}
\eea
where the last definition holds when the coefficients $(ij\ldots)$ and $(i'j'\ldots)$ are different. The factor $\delta\left( \vl{}-\vl{}\,' \right)$ is a consequence of  statistical isotropy.
Omitting terms of higher  than  fourth order in the Weyl potential and 
terms that vanish as a consequence of  Wick's theorem (odd number of Weyl potentials), we obtain 
\bea 
\tilde{C}^{\Mcal}_\ell&=& C^{{\Mcal}}_\ell+C_\ell^{(0,11)}+C_\ell^{(1, 1)}+C_\ell^{(0,13)}+C_\ell^{(0,22)}+C_\ell^{(0,112)}+C_\ell^{(0,1111)}
\nonumber \\
&  &+C_\ell^{(1, 3)}+
C_\ell^{(2, 2)}+C_\ell^{(1, 12)}+C_\ell^{(1, 111)}+C_\ell^{(2, 11)}+C_\ell^{(11, 11)}
\eea
where $C_\ell^{(0,0)}\equiv C^{{\Mcal}}_\ell$ is the unlensed power spectrum and $C_\ell^{(0,11)}$ and $C_\ell^{(1, 1)}$ are the well-known leading order corrections  given by
\bea
C_\ell^{(0,11)} &=&-C_{\ell}^\Mcal(z_s)\NINT{1}\,\left(\vl{1}\cdot \vl{}\right)^2\,
C_{\ell_1}^\psi(z_s,z_s)
\,,
\label{eq:firstorder1}
 \\
C_\ell^{(1, 1)}&=&\NINT{1}\left[\left(\vl{}-\vl{1}\right)\cdot\vl{1}\right]^2\,C_{|\vl{}-\vl{1}|}^\psi(z_s,z_s)C_{\ell_1}^\Mcal(z_s)
\,.
\label{eq:firstorder2}
\eea

The next-to-leading order corrections can be evaluated using the expressions for the deflection angle up to third order given in the previous section and inserting them in the definition of the $\Acal^{(i\cdots)}$'s 
(see Appendix \ref{a:Al}).   The expressions for these somewhat lengthy higher order corrections are also given in Appendix \ref{a:Al}.

\section{Lensed angular power spectrum: Limber approximation}
\label{Sec5}
\setcounter{equation}{0}

In order to numerically evaluate the new higher order lensing contributions to the CMB temperature anisotropies, calculated in the previous section, 
we apply the Limber approximation~\cite{Limber:1954zz,Kaiser:1991qi,LoVerde:2008re}.
 The Limber approximation is valid at large $\ell$ and,
because lensing is appreciable only for $\ell>100$, this is an excellent approximation for this paper.
  
Following~\cite{Bernardeau:2011tc}, the Limber approximation can be written as
\be 
\frac{2}{\pi}\int dk\,k^2\,f(k)
j_\ell\left(k x_1\right) j_\ell\left(k x_2\right)
\simeq \frac{\delta_D(x_1-x_2)}{x_1^2} f\left(\frac{\ell+1/2}{x_1}\right)\,,
\ee
where $f(k)$ should be a smooth, not strongly oscillating function of $k$ which decreases sufficiently rapidly for $k\ra\infty$. (More precisely, $f(k)$ has to decrease faster than $1/k$ for $k>\ell/x$.)
Using this approximation, one  obtains Limber-approximated $C_\ell$ as follows, see also~\cite{DiDio:2015bua}
 \bea
C_\ell^W(z,z')&=&
\frac{1}{2 \pi}\int dk\,k^2\,P_R(k)
\left[ T_{\Psi+\Phi}(k,z)j_\ell\left( kr \right) \right]
\left[ T_{\Psi+\Phi}(k,z')j_\ell\left(k r'\right) \right]\nonumber\\
&=&\frac{\delta\left( r'-r \right)}{r^2}\,
\frac{1}{4}P_R\left(\frac{\ell+1/2}{r}\right)
\left[T_{\Psi+\Phi}\left(\frac{\ell+1/2}{r},z\right)\right]^2 \,,
\label{CWLimber}
\eea
and, from Eq.~(\ref{Cpsi}), we find for the power spectrum of the lensing potential
\bea
C_\ell^\psi(z,z')&=&4
\int_{0}^{r}dr_1\frac{r-r_1}{r_1}
\int_{0}^{r'}dr_2\frac{r'-r_2}{r_2}\,
C_\ell^W(z_1,z_2) \nonumber\\
&=&\Theta\left( r'-r \right)\int_{0}^{r}dr_1\frac{\left(r-r_1\right)\left(r'-r_1\right)}{r\,r'\,
r_1^4}
P_R\left(\frac{\ell+1/2}{r_1}\right)
\left[T_{\Psi+\Phi}\left(\frac{\ell+1/2}{r_1},z_1\right)\right]^2\nonumber\\
&&+\,\Theta\left( r-r' \right)\int_{0}^{r'}dr_1\frac{\left(r-r_1\right)\left(r'-r_1\right)}{r\,r'\,r_1^4}
P_R\left(\frac{\ell+1/2}{r_1}\right)
\left[T_{\Psi+\Phi}\left(\frac{\ell+1/2}{r_1},z_1\right)\right]^2\,, \nonumber \\
\label{CpsiLimber}
\eea
 where $\Theta$ is the Heaviside step function.
When $z=z'$ the result~(\ref{CpsiLimber}) simplifies to 
\bea
C_\ell^\psi(z,z)&=&\int_{0}^{r}dr_1\frac{\left(r-r_1\right)^2}{r^2\,r_1^4}
P_R\left(\frac{\ell+1/2}{r_1}\right)
\left[T_{\Psi+\Phi}\left(\frac{\ell+1/2}{r_1},z_1\right)\right]^2\,.
\label{CpsiLimberSameTime}
\eea
This is in agreement with the corresponding results of Ref.~\cite{Bernardeau:2011tc}.
In~\cite{Bernardeau:2011tc} is also shown that the Limber-approximated $C_\ell^\psi$'s at equal redshift are a very good approximation 
already for $\ell>20$. We assume here that this is still true when we consider $C_\ell^\psi$ at two different redshifts.

Using the Limber-approximated $C_\ell$'s given in Eqs.~(\ref{CWLimber}) and~(\ref{CpsiLimber}) in the analytical results of the previous section, 
the expressions simplify considerably. 
As an example, let us  study $C_\ell^{(13)}$ and $C_\ell^{(22)}$.
Inserting Eq.~\eqref{CWLimber} in Eq.~\eqref{C13}, we find
\bea
C_\ell^{(0,13)}(z_s)&=&
-2\,C_{\ell}^\Mcal(z_s)
\NINT{1}\NINT{2}\left(\vl{1}\cdot\vl{}\right)\,
\left(\vl{2}\cdot\vl{}\right)\,
\ell_2^2\,
\left(\vl{2}\cdot\vl{1}\right)\nonumber\\
&&\times
\int_{0}^{r_s}dr'\frac{r_s-r'}{r_s\,r'}
\int_{0}^{r'}dr''\frac{r'-r''}{r'\,r''}\,
C_{\ell_1}^\psi(z_s,z'')\nonumber\\
&&\times
\frac{\delta\left( r'-r'' \right)}{r'^2}\,
P_R\left(\frac{\ell_2+1/2}{r'}\right)
\left[T_{\Psi+\Phi}\left(\frac{\ell_2+1/2}{r'},z'\right)\right]^2
\nonumber 
\\
& &
+2\,C_{\ell}^\Mcal(z_s)
\NINT{1}\NINT{2}
\left(\vl{2}\cdot\vl{}\right)\,
\ell_2^2\,
\left(\vl{2}\cdot\vl{1}\right)
\left(\vl{1}\cdot\vl{}\right)
\nonumber\\
&&\times
\int_{0}^{r_s}dr'\frac{r_s-r'}{r_s\,r'}
C_{\ell_1}^\psi(z_s,z')\,
\int_{0}^{r'}dr'' \frac{r'-r''}{r'\,r''}\nonumber\\
&&\times
\frac{\delta\left( r'-r'' \right)}{r'^2}\,
P_R\left(\frac{\ell_2+1/2}{r'}\right)
\left[T_{\Psi+\Phi}\left(\frac{\ell_2+1/2}{r'},z'\right)\right]^2
\nonumber 
\\
& &
+C_{\ell}^\Mcal(z_s)
\NINT{1}\NINT{2}
\left(\vl{1}\cdot\vl{}\right)^2\,
\left(\vl{1}\cdot\vl{2}\right)^2
\nonumber
\\
&&\times
\int_{0}^{r_s}dr' \frac{r_s-r'}{r_s\,r'}\,
\int_{0}^{r_s}dr''\frac{r_s-r''}{r_s\,r''}
C_{\ell_2}^\psi(z'',z'')\nonumber\\
&&\times
\frac{\delta\left( r'-r'' \right)}{r'^2}\,
P_R\left(\frac{\ell_1+1/2}{r'}\right)
\left[T_{\Psi+\Phi}\left(\frac{\ell_1+1/2}{r'},z'\right)\right]^2\,,
\label{c13LimberBS}
\eea
which simplifies to
\bea
C_\ell^{(0,13)}(z_s)
&=&C_{\ell}^\Mcal(z_s)
\NINT{1}\NINT{2}
\left(\vl{1}\cdot\vl{}\right)^2\,
\left(\vl{1}\cdot\vl{2}\right)^2
\nonumber
\\
&&\times
\int_{0}^{r_s}dr' \frac{\left(r_s-r'\right)^4}{r_s^2\,r'^4}
C_{\ell_2}^\psi(z',z')
P_R\left(\frac{\ell_1+1/2}{r'}\right)
\left[T_{\Psi+\Phi}\left(\frac{\ell_1+1/2}{r'},z'\right)\right]^2.
\label{c13Limber}
\eea
The first two terms in Eq.~(\ref{c13LimberBS}) vanish due to the $\delta(r'-r'')$ acting on the kernel $\frac{r'-r''}{r'\,r''}$. 
Similarly, using Eq.~\eqref{CWLimber} in Eq.~\eqref{C22} 
we obtain
\bea
C_\ell^{(0,22)}(z_s) 
&=&-
C_{\ell}^\Mcal(z_s)
\NINT{1}\NINT{2}\left(\vl{}\cdot\vl{1}\right)^2\,
\left(\vl{1}\cdot\vl{2}\right)^2
\nonumber
\\
&&\times
\int_{0}^{r_s}dr'\frac{\left(r_s-r'\right)^2}{r_s^2\,r'^4}
C_{\ell_2}^\psi(z',z')
P_R\left(\frac{\ell_1+1/2}{r'}\right)
\left[T_{\Psi+\Phi}\left(\frac{\ell_1+1/2}{r'},z'\right)\right]^2.
\eea
It is easy to check that $C_\ell^{(0,22)}\equiv -C_\ell^{(0,13)}$ within the Limber approximation.  

Similar cancellations occur when using 
Eq.~\eqref{CWLimber} to evaluate the terms  
$C_\ell^{(0,1111)}$, $C_\ell^{(1, 3)}$, $C_\ell^{(2, 2)}$, $C_\ell^{(1, 111)}$,  and $C_\ell^{(11, 11)}$. These cancellations are most probably the result of a consistency relation: large angle inhomogeneities cannot affect the spectrum on small angles as they just correspond to an isotropic Universe at a slightly different temperature. Some of the cancellations have also been found in~\cite{Pratten:2016dsm}. For the same reason, also the terms in $C_\ell^{(0,112)}$ cancel perfectly within the Limber approximation.
Furthermore, because of the Limber approximation, we can always use 
Eqs.~(\ref{CpsiLimber}) and~(\ref{CpsiLimberSameTime}) when evaluating the power spectrum of the lensing potential.
All the remaining non-vanishing $C_\ell^{(0,ijk...)}$ and $C_\ell^{(ij,i'j')}$ are presented in the next section.

\section{Lensed angular power spectrum: final results}
\label{Sec6}
\setcounter{equation}{0}
In order to interpret properly corrections from different terms, let us classify the higher order non-null 
$C_\ell^{(...)}$ 
in three different groups.

\subsection{First group}
\label{num-res-s1}
In this first group we collect the next-to-leading order contributions only contain first order deflection angles  $\theta^{a(1)}$. 
These are given by
\bea
C_\ell^{(0,1111)}\!&=&
\frac{1}{4}\frac{\left( C_\ell^{(0,11)} \right)^2}{C_\ell^\Mcal}\,,
\label{GroupOne1}\\
&&
\text{coming from } \langle \th{1}{a}\th{1}{b}\th{1}{c}\th{1}{d} \rangle
\langle \nabla_a\nabla_b\nabla_c\nabla_d\Mcal \bar\Mcal \rangle \,,\nonumber\\
C_\ell^{(1, 111)}\!&=&\!
-\!\!\!\NINT{1}\!\!\NINT{2}\!\left[\left(\vl{}-\vl{1}\right)\cdot\vl{1}\right]^2\left(\vl{2}\cdot\vl{1}\right)^2\,
\!C_{\ell_1}^\Mcal(z_s)
C_{|\vl{}-\vl{1}|}^\psi(z_s,z_s)C_{\ell_2}^\psi(z_s,z_s),
\label{GroupOne2}\\
&&\text{coming from  } \langle \th{1}{a}\th{1}{b}\th{1}{c}\th{1}{d} \rangle
\langle \nabla_a\Mcal \nabla_b\nabla_c\nabla_d\bar\Mcal \rangle \,,\nonumber\\
C_{\ell_1}^{(11, 11)}\!&=&\frac{1}{4}\frac{\left(C_\ell^{(0,11)}\right)^2}{C_\ell^\Mcal}
+\frac{1}{2}\NINT{1}\NINT{2}\,\left\{\left[\left(\vl{}-\vl{1}+\vl{2}\right)\cdot \vl{1}\right]\,
\left(\vl{1}\cdot \vl{2}\right)\right\}^2\,
C_{\ell_1}^\Mcal(z_s)\nonumber\\
&&\times C_{|\vl{}-\vl{1}+\vl{2}|}^\psi(z_s,z_s)C_{\ell_
2}^\psi(z_s,z_s)\,,
\label{GroupOne3}\\
&&\text{coming from  } \langle \th{1}{a}\th{1}{b}\th{1}{c}\th{1}{d} \rangle
\langle \nabla_a\nabla_b\Mcal\nabla_c\nabla_d\bar\Mcal \rangle \,.
\nonumber
\eea

\begin{figure}[ht!]
\centering
\includegraphics{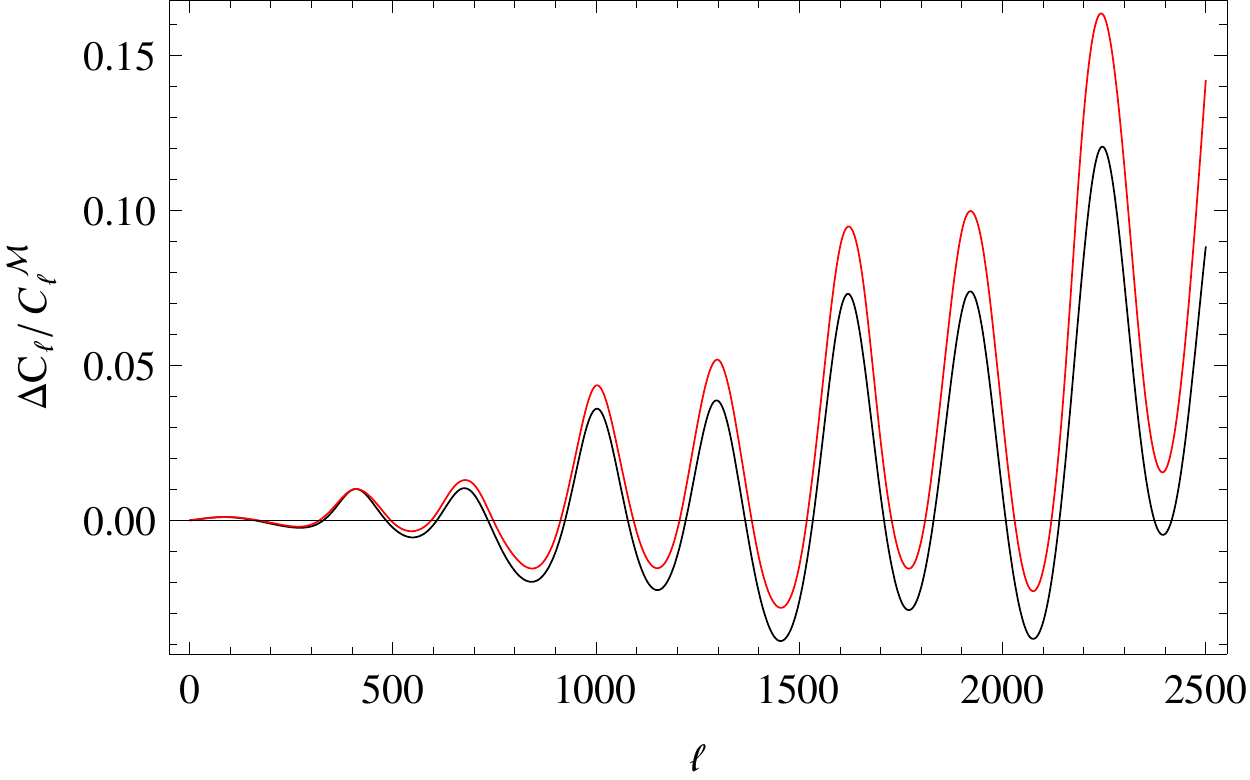}
\caption{Corrections due to $\th{1}{a}$. The red curve refers to first order terms given in Eqs.~\eqref{eq:firstorder1} and \eqref{eq:firstorder2}, $\Delta C_\ell=C_\ell^{(11)}+C_\ell^{(1, 1)}$, while black curve takes into account  the fully re-summed first order contribution $\Delta C_\ell=\tilde C^{\Mcal\, (1)}_\ell-C^{\Mcal}_\ell$. 
This result is well known and can be found also e.g. in~\cite{Lewis:2006fu}.}
\label{CompFirstTerm-Resum-Order1}
\end{figure}

More generally, corrections due to the first order deflection angles can be considered beyond the small deflection angle approximation in a non-perturbative sense~\cite{Seljak:1995ve,RuthBook,Lewis:2006fu}. 
Indeed, by using the Gaussianity of $\theta^{a(1)}$, exponentiation allows to sum all the corrections coming only from the first order deflection angles  non-perturbatively. 
This approach is based on the property that a Gaussian stochastic variable $y$ is completely determined by its 2-point statistics, and it is easy to verify that for a Gaussian variable $\langle e^{i y} \rangle=e^{-\langle y^2 \rangle/2}$. 
Hence, the correction to the correlation function $\xi(r)$ due to lensing from a Gaussian $\delta\boldsymbol{\theta}$ can be  taken into account  as follows:
\bea
\tilde\xi(r)&=&\langle \tilde\Mcal({\bf x}) \tilde\Mcal({\bf x}+{\bf r}) \rangle=
\langle \Mcal({\bf x}+\delta \boldsymbol{\theta}) \Mcal({\bf x}+{\bf r}+\delta \boldsymbol{\theta}') \rangle
\nonumber
\\
&=&\NINT{}\,C_\ell^\Mcal e^{i\vl{}\cdot{\bf r}}\langle e^{i\vl{}\cdot\left( \delta \boldsymbol{\theta}- \delta \boldsymbol{\theta}' \right)} \rangle = \NINT{}\,C_\ell^\Mcal e^{i\vl{}\cdot  {\bf r}} e^{- \langle \left[ \vl{}\cdot\left(\delta\boldsymbol{\theta}-\delta\boldsymbol{\theta}' \right) \right]^2 \rangle /2} \,.
\label{eq:exp}
\eea
In the approach used in present public CMB codes like \camb{}\footnote{\url{http://camb.info}}~\cite{Lewis:1999bs} and \class{}\footnote{\url{http://class-code.net}}~\cite{Lesgourgues:2011re,Blas:2011rf}, to compute the expectation value
$\langle \left[ \vl{}\cdot\left(\delta \boldsymbol{\theta}-\delta\boldsymbol{\theta}' \right) \right]^2 \rangle$ one 
assume $\delta  \boldsymbol{\theta}=\boldsymbol{\theta}^{(1)}$ and
defines the matrix
\bea
A_{a b} \left( {\bf r} \right) = \langle \theta_a^{(1)} \left( {\bf x} \right) \theta_b^{(1)} \left( {\bf x} + {\bf r} \right) \rangle = \int \frac{d^2 \ell}{\left( 2 \pi \right)^2} \ell_a \ell_b C^\psi_\ell e^{i {\bf r} \cdot \boldsymbol{\ell}}\,,
\eea
where we have used Eqs.~(\ref{Weylpot}) and (\ref{TheOrd1}). 
Statistical isotropy constrains  the matrix $(A_{ab})$ to be of the form 
\be
A_{a b} \left( {\bf r} \right) = \frac{1}{2} A_0 \left( r \right) \delta_{ab} - A_2 \left( r \right) \left( \frac{r_a r_b}{r^2} - \frac{1}{2} \delta_{ab} \right)\,,
\ee
with
\be
A_0 \left( r\right) = \int \frac{d \ell \ \ell^3}{ 2 \pi } C^\psi_\ell J_0 \left( r \ell \right)
\ee
and
\be
A_2 \left( r\right) = \int \frac{d \ell \ \ell^3}{ 2 \pi } C^\psi_\ell J_2 \left( r \ell \right) \, .
\ee
Here $J_0$ and $J_2$ are the Bessel functions of order zero and two. Therefore we find
\be
\langle \left[ \vl{}\cdot\left(\delta \boldsymbol{\theta}-\delta\boldsymbol{\theta}' \right) \right]^2 \rangle = \ell^2 \left( A_0 \left( 0 \right)  - A_0 \left( r \right) +A_2 \left( r \right) \cos \left( 2 \phi \right) \right)\,,
\ee
where $\cos \left(  \phi \right) = \hat{ {\bf r}} \cdot \hat { \boldsymbol{\ell}}$. With this, the lensed power spectrum finally becomes
\be 
\tilde C^{\Mcal\, (1)}_\ell = \int dr  r J_0 \left( \ell r \right) \int \frac{d^2 \ell'}{ \left( 2 \pi \right)^2} C^{\Mcal}_{\ell'} e^{-i \boldsymbol{\ell}' \cdot {\bf r}} \exp \left[ -\frac{\ell'^2}{2} \left( A_0 \left( 0 \right) - A_0 \left( r \right) + A_2 \left( r \right) \cos \left( 2 \phi \right) \right) \right] \,.
\label{ExpFirstOrder}
\ee
The solution~(\ref{ExpFirstOrder})  captures the full correction, from first order deflection angles alone, to the unlensed $C_\ell^{\Mcal}$ for arbitrary 
large $\ell$, since the first order perturbation angle is fully re-summed \cite{Seljak:1995ve,RuthBook,Lewis:2006fu}. On the other hand, as pointed out before, since deflections angles are typically of the order of arc minutes, the small deflection 
angle approximation is sufficient for about $\ell\lesssim 2500$. This is also  shown in Fig.~\ref{CompFirstTerm-Resum-Order1}, 
where we compare the correction on the unlensed $C_\ell^{\Mcal}$ obtained 
considering only the leading correction in the small-deflection angle approximation given in Eqs.~\eqref{eq:firstorder1} and~\eqref{eq:firstorder2} with the correction obtained taking into account the re-summed contributions considering only $\th{1}{a}$ (see Eq.~\eqref{ExpFirstOrder}).
 As we see from this figure, for $\ell\gtrsim 1500$ the correction from the resummed lensing, Eq.~(\ref{ExpFirstOrder}), is about $25\%$ smaller than the one obtained at lowest order, Eqs.~(\ref{eq:firstorder1}) and~(\ref{eq:firstorder2}),
while the difference is even smaller for  lower multipoles $\ell$. In general, we observe that keeping only the first term of the sum,
 which corresponds to  the leading term of the small-deflection angle approximation,  we obtain always the right 
 order of magnitude of the correction for $\ell\lesssim 2500$, and it is a good approximation at even smaller 
 $\ell$. 
 Since the higher order deflections angles are so much smaller, we expect the lowest order approximation to be significantly better at higher order.
  
We stress again that the exact relation between the lensing potential $C^\psi_\ell$ and the lensed CMB power spectrum of Eq.~(\ref{ExpFirstOrder}) 
takes into account only the non-perturbative contribution coming from the first order deflection angles, and  it is valid because $\th{1}{a}$ has  Gaussian statistic.

Going to higher order, taking into account post-Born corrections, generates further terms which lead to non-Gaussian contributions to the deflection angle. In Ref.~\cite{Hagstotz:2014qea} post-Born corrections have been studied only via
a generalization of Eq.~(\ref{ExpFirstOrder}),
which takes into account both the higher order correction to the power spectrum of lensing potential $C_\ell^{\varPsi \varPsi}$ and of the curl potential $C_\ell^{\Omega \Omega}$. 
This generalization is still based on the assumption that $\delta \boldsymbol{\theta}$
has a Gaussian statistic and uses  the relation 
$\langle e^{i y} \rangle=e^{-\langle y^2 \rangle/2}$ for the full non-Gaussian deflection angle. This is obviously not correct  beyond 
linear order in  the deflection angle.
One consequence of using the generalization of Eq.~(\ref{ExpFirstOrder}) to account for  the post-Born corrections to 
$C_\ell^{\Mcal}$  
is that only contributions with even power of $\delta \boldsymbol{\theta}$ are considered.
Contributions related to the non-Gaussian statistics of 
$\delta \boldsymbol{\theta}$, which appear in the terms
with an odd number of deflection angles generated by Taylor expanding  $\langle e^{i\vl{}\cdot\left( \delta \boldsymbol{\theta}- \delta \boldsymbol{\theta}' \right)} \rangle$,   are neglected.
 
In the following, we divide our leading post-Born corrections, which are not taken into account in the first group, into a second and a third group. The second group corresponds to the  terms that are present also if $\delta \boldsymbol{\theta}$ would have a Gaussian statistics beyond linear order.
These  would be the first terms of the tentative re-summation performed in~\cite{Hagstotz:2014qea}. 
While the third group contains the leading terms (in the perturbative number of gravitational potentials) of the 
non-Gaussian post-Born correction, which contain an odd number of deflection angles and are therefore not  considered in the re-summation of Ref.~\cite{Hagstotz:2014qea}, and in general in previous research.

Following the result shown in Fig.~\ref{CompFirstTerm-Resum-Order1} for the corrections related only to $\theta^{(1)}$, we assume that the small-deflection angle approximation holds also for the post-Born higher order corrections that we determine in the following up to $\ell\simeq 2500$.  Since higher order corrections are smaller than the first order ones, we believe that this  is a safe assumption.

\subsection{Second group}
\label{num-res-s2}

In this second group we study the leading post-Born corrections, which come from
the deflection angles up to third order when these appear in pairs like 
$\langle \th{2}{a}\th{2}{b} \rangle$ and $\langle \th{1}{a}\th{3}{b} \rangle$.
They are given by
\bea
C_\ell^{(1, 3)} &=&-\NINT{1}\NINT{2}
\left[\left(\vl{}-\vl{1}\right)\cdot\vl{1}\right]^2\,
\left[\left(\vl{}-\vl{1}\right)\cdot\vl{2}\right]^2\,
C_{\ell_1}^\Mcal(z_s)
\nonumber\\
&&\times
\int_{0}^{r_s}dr'
\frac{\left(r_s-r'\right)^2}{r_s^2\,r'^4}
C_{\ell_2}^\psi(z',z')
\nonumber\\
&&\times
P_R\left(\frac{|\vl{}-\vl{1}|+1/2}{r'}\right)
\left[T_{\Psi+\Phi}\left(\frac{|\vl{}-\vl{1}|+1/2}{r'},z'\right)\right]^2\,,\label{eq:secondgroup1}\\
&& \text{coming from }~~ \langle \th{1}{a}\th{3}{b} \rangle
\langle \nabla_a\Mcal \nabla_b\bar\Mcal \rangle\,,
\nonumber
\\
C_\ell^{(2, 2)} &=&
\NINT{1}\NINT{2}\,\left[\left(\vl{}-\vl{1}+\vl{2}\right)\cdot\vl{1}\right]^2\,
\left[\left(\vl{}-\vl{1}+\vl{2}\right)\cdot\vl{2}\right]^2\,C_{\ell_1}^\Mcal(z_s)
\nonumber\\
&&\times\int_{0}^{r_S}dr'\frac{\left(r_s-r'\right)^2}{r_s^2\,r'^4}C_{\ell_2}^\psi(z',z')
\nonumber\\
&&\times P_R\left(\frac{|\vl{}-\vl{1}+\vl{2}|+1/2}{r'}\right)
\left[T_{\Psi+\Phi}\left(\frac{|\vl{}-\vl{1}+\vl{2}|+1/2}{r'},z'\right)\right]^2\,,
\label{eq:secondgroup2}\\
&& \text{coming from } ~~\langle \th{2}{a}\th{2}{b} \rangle
\langle \nabla_a\Mcal \nabla_b\bar\Mcal \rangle\,.
\nonumber
\eea
Contrary to $C_\ell^{(0,13)}$ and $C_\ell^{(0,22)}$, the two contributions above
do not exactly cancel.
On the other hand,
in the range of integration for which $|\vl{}-\vl{1}+\vl{2}|\simeq
|\vl{}-\vl{1}|$, i.e. when $\ell_2$ is small,  the two above integrands are nearly identical. This leads to a significant but not exact
cancellation between
$C_\ell^{(1,3)}$ and $C_\ell^{(2,2)}$.

The physical interpretation of this is that for $|\vl{}-\vl{1}|\gg\ell_2$ a deflection angle related to the $\ell_2$  acts like a 'global rotation' which has no effect on the lensing corrections to the CMB power spectrum. Therefore, all the contributions from $|\vl{}-\vl{1}|\gg\ell_2$  cancel. This can be considered as a 'consistency relation' for the lensing corrections to the CMB power spectrum.
Similar cancellations have also been found in~\cite{Pratten:2016dsm}.

\begin{figure}[h!]
\centering
\includegraphics[scale=0.6]{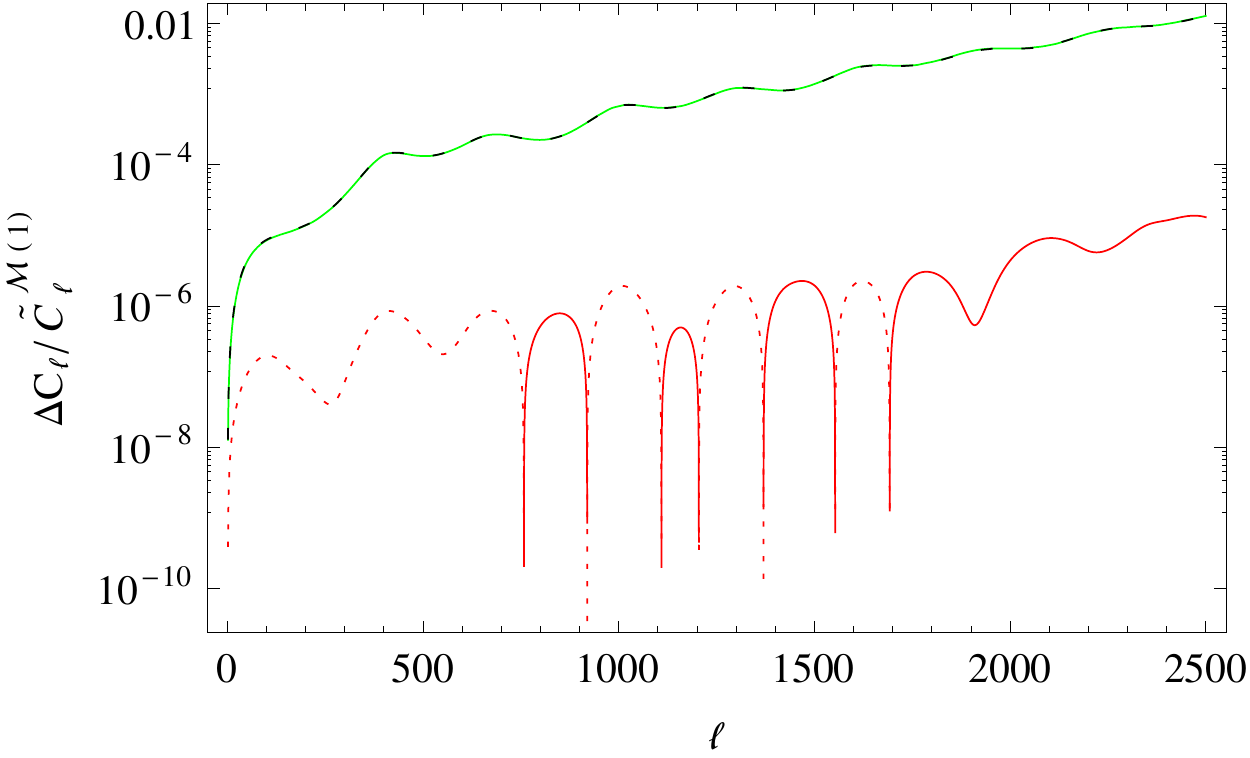}
\includegraphics[scale=0.6]{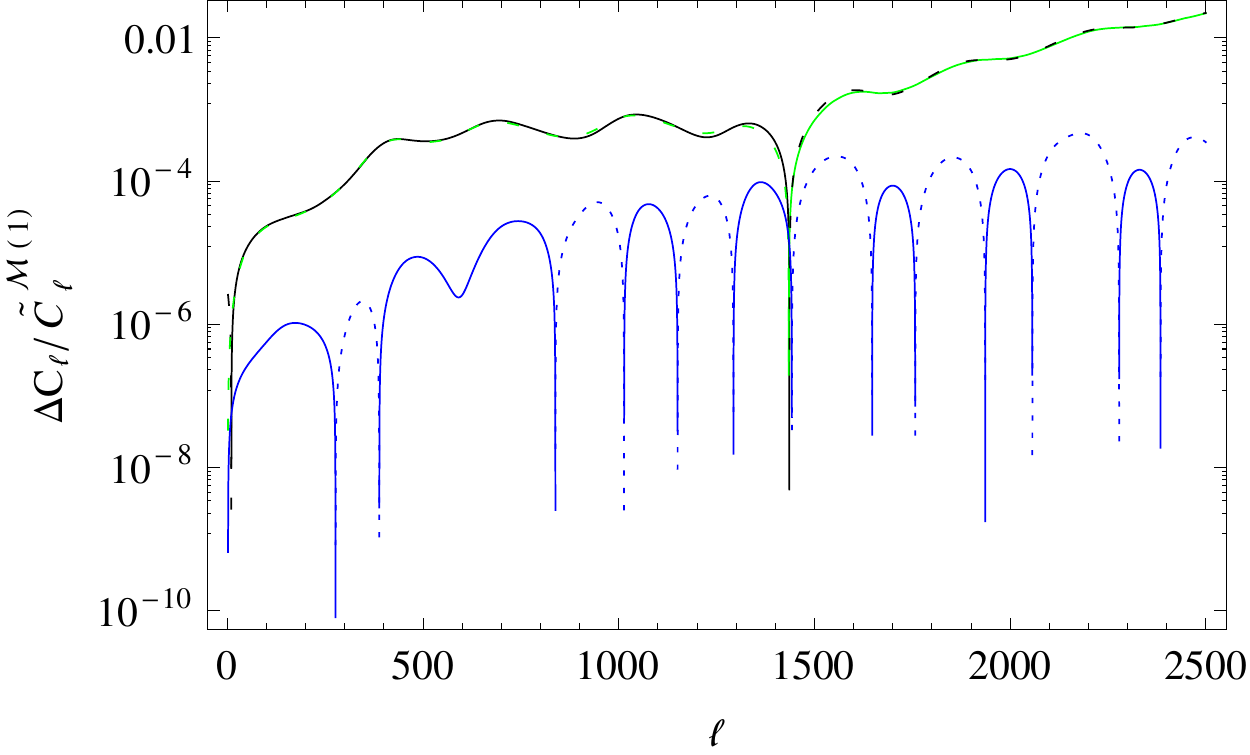}
\caption{Left panel: we show the partial cancellation of the second group.
$\Delta C_\ell=C_\ell^{(1,3)}$ (black curve),
$\Delta C_\ell=C_\ell^{(2,2)}$  (green curve) and the sum of them (red curve). Negative parts are dashed. 
The cancellation reduces the final result by about a factor of 700 for $\ell \sim 2500$.
Right panel:  we show the partial cancellation of the third group.  $\Delta C_\ell=C^{(1,12)}$ (black curve),
$\Delta C_\ell=C^{(2,11)}$ (green curve), 
and the sum of them (blue curve). Negative parts are dashed. 
The cancellation reduces the final result by about a factor of 50 for $\ell \sim 2500$.}
\label{f:cancel}
\end{figure}

Actually, each term appearing in the lensing corrections has a 'counter term' which exactly cancels it in this limit. In Fig.~\ref{f:cancel} we show two examples of this partial cancellation which reduces the final result at large $\ell$ up to  two orders of magnitude. 

\subsection{Third group}
\label{num-res-s3}

In this third group we consider terms with three deflection angles which do not vanish due to the non-Gaussian statistic of $\theta^{a(2)}$.
These are the following two contributions 
\bea
C_\ell^{(1, 12)} 
&=&
-2\,\NINT{1}\NINT{2}
\left(\vl{1}\cdot\vl{2}\right)\,
\left[(\vl{}-\vl{1})\cdot\vl{2}\right]\,
\left[\left(\vl{}-\vl{1}\right)\cdot\vl{1}\right]^2\nonumber\\
&&\times\,C_{\ell_1}^\Mcal(z_s)
\int_{0}^{r_s}dr'\frac{\left(r_s-r'\right)^2}{r_s^2\,r'^4}\,
P_R\left(\frac{|\vl{}-\vl{1}|+1/2}{r'}\right)\nonumber\\
&&\times\left[T_{\Psi+\Phi}\left(\frac{|\vl{}-\vl{1}|+1/2}{r'},z'\right)\right]^2
C_{\ell_2}^\psi\left( z_s,z' \right)\,,\label{eq:thirdgroup1}\\
&&\text{coming from } ~~ \langle \th{1}{a}\th{1}{b}\th{2}{c} \rangle
\langle \nabla_a\Mcal \nabla_b\nabla_c\bar\Mcal \rangle\,,
\nonumber
\\
C_\ell^{(2, 11)} 
&=&
2\,\NINT{1}\NINT{2}\,
\left(\vl{1}\cdot \vl{2}\right)\,
\left[\left(\vl{}-\vl{1}+\vl{2}\right)\cdot\vl{2}\right]
\left[\left(\vl{}-\vl{1}+\vl{2}\right)\cdot \vl{1}\right]^2\,
\nonumber\\
&&\times\,C_{\ell_1}^\Mcal(z_s)
\int_{0}^{r_s}dr'\frac{\left(r_s-r'\right)^2}{r_s^2\,r'^4}
P_R\left(\frac{|\vl{}-\vl{1}+\vl{2}|+1/2}{r'}\right)\nonumber\\
&&\times
\left[T_{\Psi+\Phi}\left(\frac{|\vl{}-\vl{1}+\vl{2}|+1/2}{r'},z'\right)\right]^2
C_{\ell_2}^\psi\left( z_s,z' \right)\,,
\label{eq:thirdgroup2}\\
&&\text{coming from } ~~ \langle \th{2}{a}\th{1}{b}\th{1}{c} \rangle
\langle \nabla_a\Mcal \nabla_b\nabla_c\bar\Mcal \rangle \,.
\nonumber
\eea
We note that,  also in this case, in the range of integration for which $|\vl{}-\vl{1}+\vl{2}|\simeq
|\vl{}-\vl{1}|$ the two above integrands are nearly identical and the corresponding contributions to
$C_\ell^{(1,12)}$ and $C_\ell^{(2,11)}$ partially cancel.  

The physical explanation of this cancelation is the same as in the second group. 
However, in this group the integral from the domain where cancellation happens, $|\vl{}-\vl{1}| \gg \ell_2$ contributes less and the cancellation is more than an order of magnitude less significant than for the second group. This is clearly visible in Fig.~\ref{f:cancel} above.

It is interesting to notice that the terms included in the second group do contribute to the lensing and curl potentials in the post-Born approximation, and they are  considered already in~\cite{Hagstotz:2014qea} (modulo a possible sign error), but the terms of the third group do not appear in these potentials, they  are 
new contributions.

We can now write the fully corrected lensed CMB temperature anisotropy power spectrum in the form
\be
\tilde C_\ell^{\Mcal} =\tilde C_\ell^{\Mcal\, (1)} + \Delta C_\ell^{(2)} +  \Delta C_\ell^{(3)}  \label{eq:cltot}
\ee
where
\bea
 \Delta C_\ell^{(2)} &=& C_\ell^{(1, 3)} + C_\ell^{(2, 2)}  \label{eq:cl2}\\
 \Delta C_\ell^{(3)} &=& C_\ell^{(1, 12)} + C_\ell^{(2, 11)}  \label{eq:cl3}  \,.
\eea
Here the $\tilde C_\ell^{\Mcal\, (1)} $ term denotes the well known resummed correction from the first order
deflection angle given in Eq.~\eqref{ExpFirstOrder}, while  $\Delta C_\ell^{(2)}$ and  $\Delta C_\ell^{(3)}$ denote the post-Born corrections from the second and third group respectively.

\subsection{Numerical Results}
\label{num-res-s4}

In this subsection we show the numerical evaluation of the  results given  above, considering both linear and non-linear (Halofit model \cite{Smith:2002dz,Takahashi:2012em}) power 
spectra for the gravitational potential. 
More precisely, here and above  the figures have been generated with the following cosmological parameters $h = 0.67$, $\omega_\text{cdm}=0.12$, $\omega_b = 0.022$ and vanishing curvature. The primordial curvature power spectrum has the amplitude $A_s = 2.215 \times 10^{-9}$, the pivot scale $k_\text{pivot} = 0.05 \ \text{Mpc}^{-1}$, the spectral index $n_s = 0.96$ and no running, compatible with~\cite{Planck:2015xua}. The Bardeen transfer function $T_{\Phi+\Psi}$ has been computed with~\class{}~\cite{Blas:2011rf}, using Halofit \cite{Takahashi:2012em} for the non-linear case.

\begin{figure}[h!]
\centering
\includegraphics[scale=1.1]{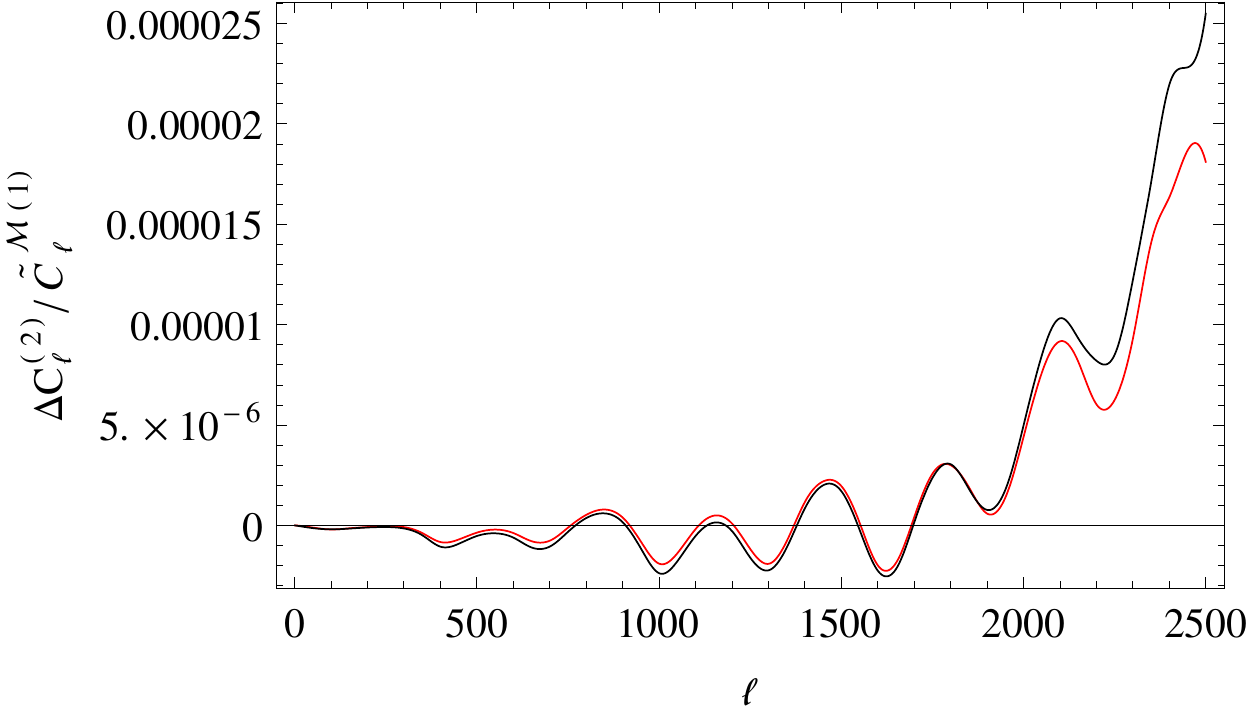}
\caption{The correction with respect to the standard lensed temperature power spectrum
$\tilde C^{\Mcal (1)}_\ell$, given in Eq.(\ref{ExpFirstOrder}),
from the second group of terms $\Delta C_\ell^{(2)}$ (see Eq.~\eqref{eq:cl2}), for linear (red curve) and non-linear (black curve) power spectra.}
\label{G-figure}
\end{figure}

In Fig.~\ref{G-figure} we plot the relative
correction $\Delta C_\ell^{(2)}/\tilde C^{\Mcal (1)}_\ell$ to the lensed temperature anisotropy spectrum given in Eq.~(\ref{ExpFirstOrder}) obtained when considering the 
post-Born correction of the second group, Eqs.~\eqref{eq:secondgroup1} and~\eqref{eq:secondgroup2}, assuming  linear and non-linear  power spectra. This correction is the post-Born contribution,
coming from deflection angles up to third order, that is present in a generalization of Eq.~(\ref{ExpFirstOrder}) as
performed in~\cite{Hagstotz:2014qea}. As a consequence, this is the effect that should be compared with the result given in~\cite{Hagstotz:2014qea}. 
This comparison shows how in \cite{Hagstotz:2014qea} the corresponding contribution is overestimated by two orders of magnitude. This overestimation is probably due to the sign error pointed out in 
Section \ref{Sec2}.

\begin{figure}[ht!]
\centering
\includegraphics[scale=1.1]{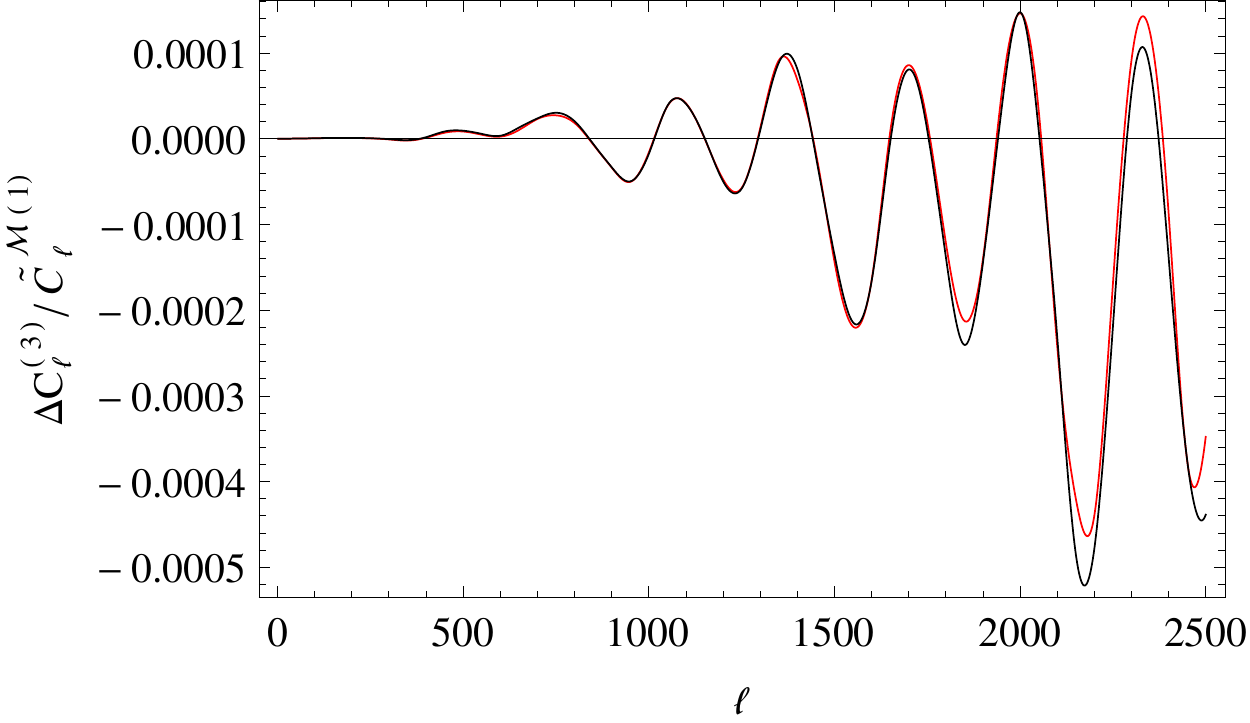}
\caption{
The correction with respect to the standard lensed temperature power spectrum
$\tilde C^{ \Mcal (1)}_\ell$, given in Eq.(\ref{ExpFirstOrder}),
from the third group of terms $\Delta C_\ell^{(3)}$ (see Eq.~\eqref{eq:cl3}), for linear (red curve) and non-linear (black curve) power spectrum.
This correction is the first non-Gaussian contribution
coming from $\th{2}{a}$.
}
\label{NG-figure}
\end{figure}

In Fig.~\ref{NG-figure} we plot the  relative
correction $\Delta C_\ell^{(3)}/\tilde C^{\Mcal (1)}_\ell$ to the lensed temperature anisotropy spectrum given in Eq.~(\ref{ExpFirstOrder}) 
obtained when we consider the 
post-Born correction from the third group, Eqs.~\eqref{eq:thirdgroup1} and \eqref{eq:thirdgroup2}, both for linear and non-linear power spectra. This correction is the first non-Gaussian contribution
coming from $\th{2}{a}$.
This contribution is not considered in~\cite{Hagstotz:2014qea}, and in general when the next-to-leading order corrections are evaluated via the amplification matrix and using the relation 
$\langle e^{i y} \rangle=e^{-\langle y^2 \rangle/2}$. 
 But, as can be seen comparing Figs.~\ref{G-figure} and~\ref{NG-figure}, this contribution is 
more than one order of magnitude larger than the total contribution from the second group.  
 Each individual term of this group is about a factor of two larger than the terms of the second group. 
But more importantly, the cancellation is more than one order of magnitude less pronounced at argued above.

Figs.~\ref{G-figure} and, especially, Fig.~\ref{NG-figure} are the main results of this paper.
Our new contribution shown in Fig.~\ref{NG-figure}, which is neglected in~\cite{Hagstotz:2014qea},  dominates
the terms of the second group by more of one order of magnitude.

\begin{figure}[ht!]
\centering
\includegraphics{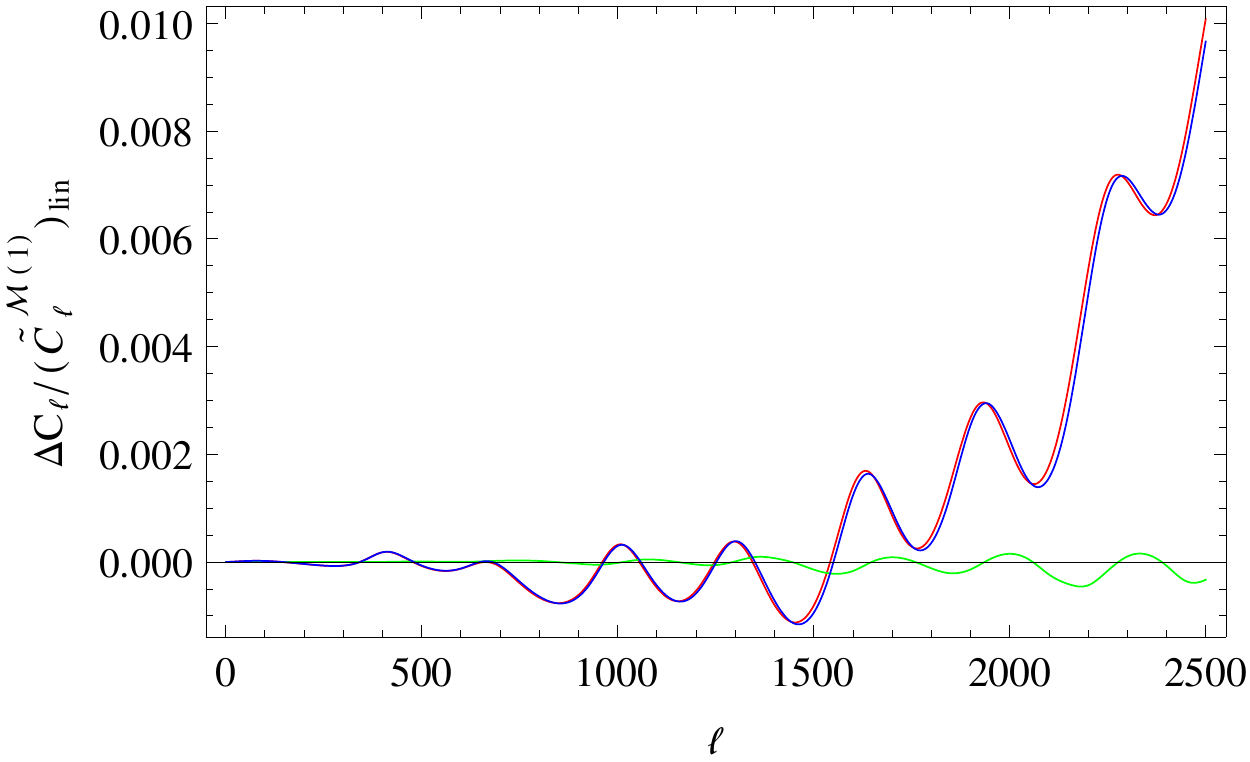}
\caption{Correction to the lensed $\tilde C^{\Mcal (1)}_\ell$, evaluated with a linear power spectrum and given in Eq.~(\ref{ExpFirstOrder}), due to the non-linear matter power spectrum (Halofit model, red curve), the  terms beyond the Born approximation ($\Delta C_\ell=\Delta C_\ell^{(2)}+\Delta C_\ell^{(3)}$, green curve) and both  (blue curve). }
\label{All-effects}
\end{figure}

The full result is presented in Fig.~\ref{All-effects}.
 In this figure we show the corrections to the lensed power spectrum 
 $\tilde C^{\Mcal (1)}_\ell$ from the nonlinearities of the matter power spectrum, we plot
  $$\left((\tilde C^{\Mcal (1)}_\ell)_{\rm non-lin} - (\tilde C^{\Mcal\, (1)}_\ell)_{\rm lin}\right)/ (\tilde C^{\Mcal (1)}_\ell)_{\rm lin}\,,
 $$
 from the  post-Born corrections, we plot
 $$
 \left((\Delta C_\ell^{(2)})_{\rm lin} + (\Delta C_\ell^{(3)})_{\rm lin}\right)/ (\tilde C^{\Mcal (1)}_\ell)_{\rm lin}\,,
 $$
 and from both post-Born corrections and a non-linear power spectrum, we plot 
 $$
 \left((\tilde C^{\Mcal (1)}_\ell)_{\rm non-lin} +(\Delta C_\ell^{(2)})_{\rm non-lin} + (\Delta C_\ell^{(3)})_{\rm non-lin}- (\tilde C^{\Mcal\, (1)}_\ell)_{\rm lin}\right)/ (\tilde C^{\Mcal (1)}_\ell)_{\rm lin}\,.
 $$
  As one can see, including a non-linear power spectrum is much more important than the post-Born correction. The post-Born correction increases from $0.01\%$  at $\ell \sim 1000$ to $0.05\%$ for $\ell \simeq 2500$, while the non-linearities give corrections of up to $1\%$ at $\ell=2500$. 

\begin{figure}[ht!]
\centering
\includegraphics[scale=0.6]{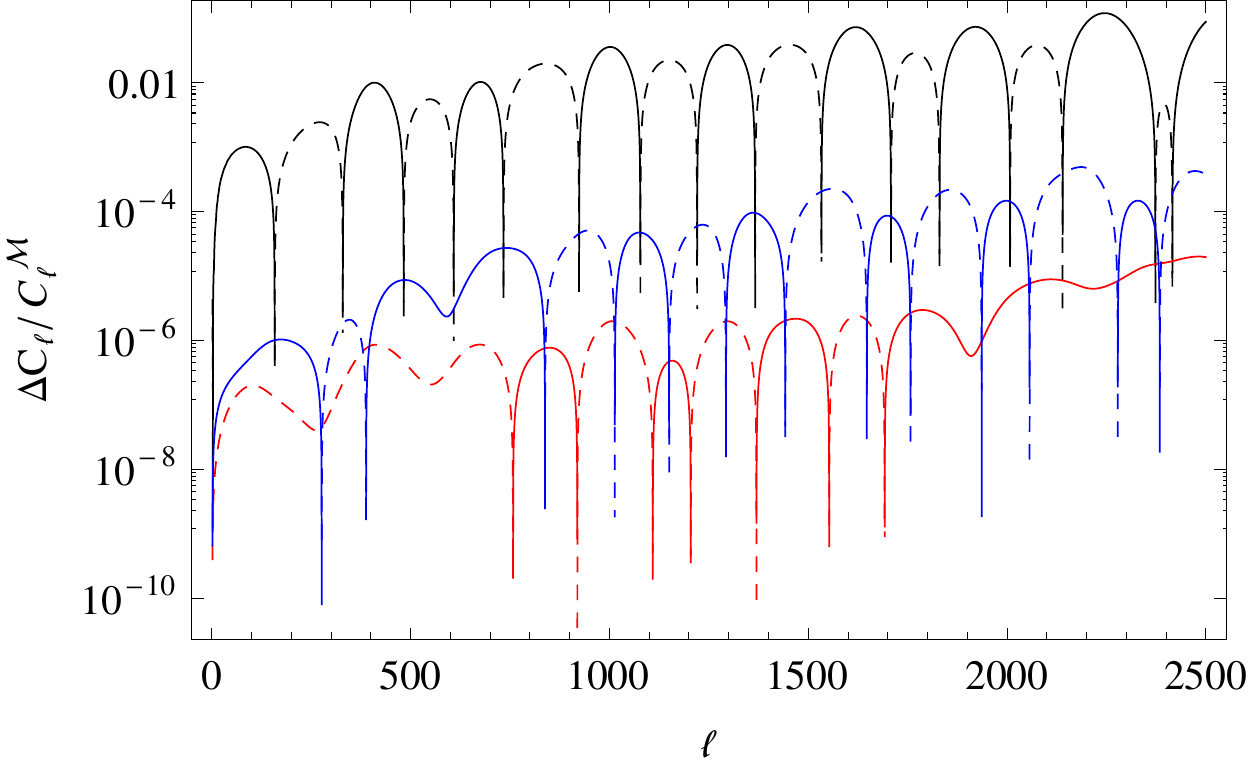}
\includegraphics[scale=0.6]{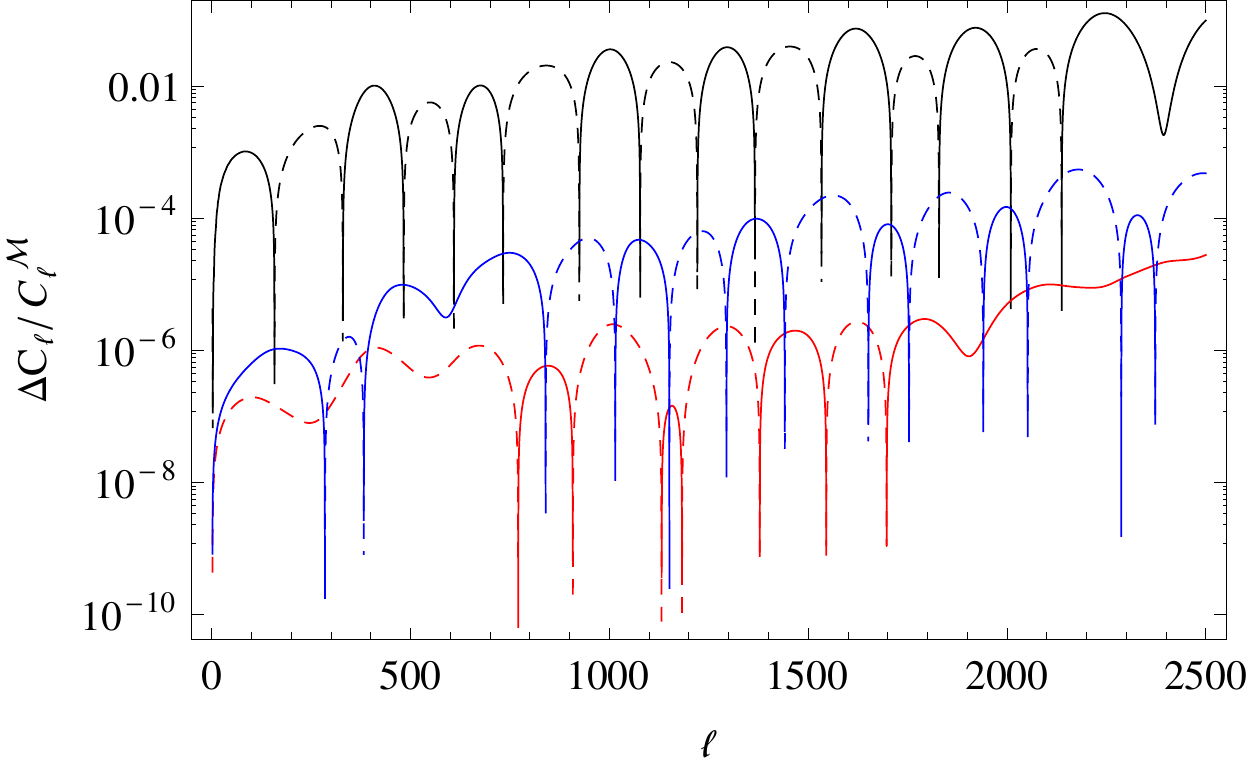}
\caption{Corrections to the unlensed $C_\ell^\Mcal$'s due to resummed $\th{1}{a}$ ($\Delta C_\ell=\tilde C^{\Mcal\, (1)}_\ell-C^{\Mcal}_\ell$, black curve), the second group ($\Delta C_\ell=
\Delta C_\ell^{(2)}$, red curve) and the third group ($\Delta C_\ell=
\Delta C_\ell^{(3)}$, blue curve) for the linear matter power spectrum (left panel) and the Halofit model (right panel). Dashed lines are negative parts. 
As we can see from these figures, the non-Gaussian part is the most relevant 
post-Born correction in going beyond the first order deflection angles contribution, for
both linear and Halofit models.}
\label{Comparing-all}
\end{figure}

\begin{figure}[ht!]
\centering
\includegraphics{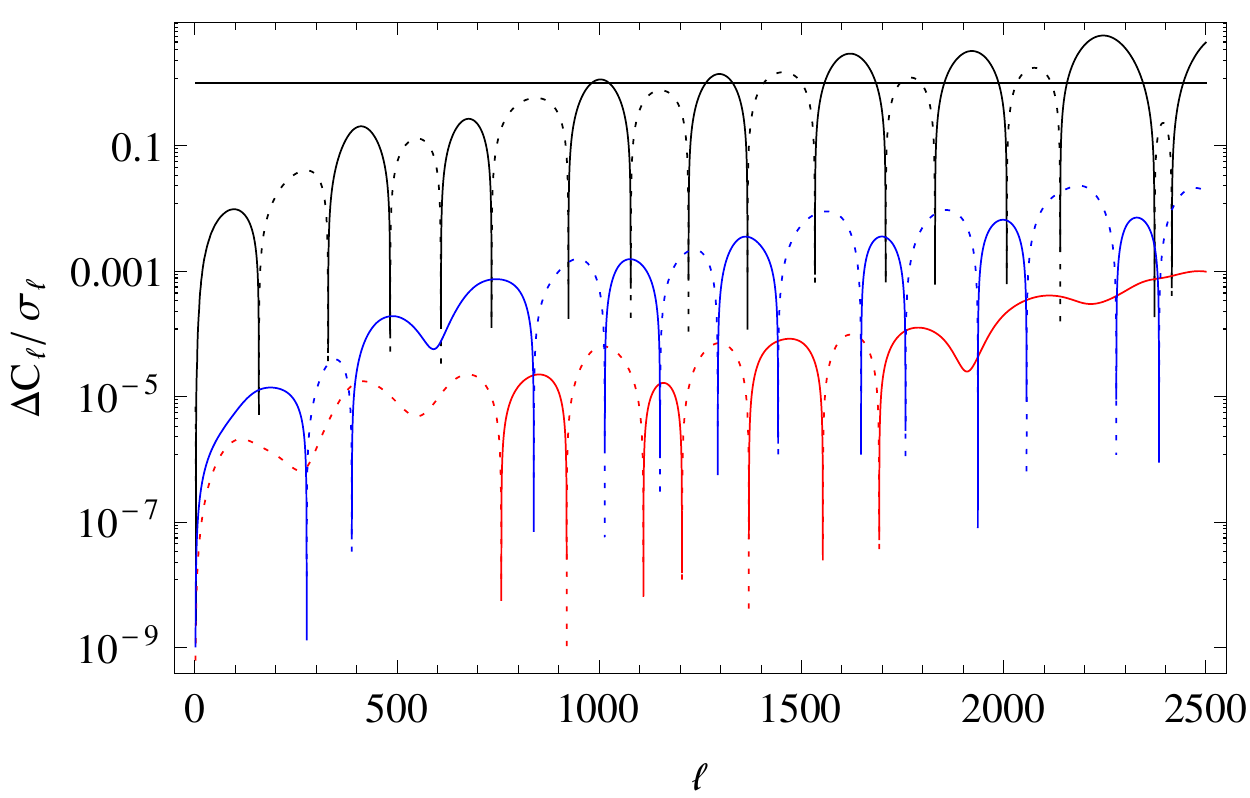}
\caption{The ratios of the lensing corrections and the cosmic variance $\sigma_\ell\equiv\sqrt{\frac{2}{2\ell+1}}C^{\Mcal}_\ell$ for the resummed $\th{1}{a}$ ($\Delta C_\ell=\tilde C^{\Mcal\, (1)}_\ell-C^{\Mcal}_\ell$, black curve), the second group ($\Delta C_\ell=
\Delta C_\ell^{(2)}$, red curve) and the third group ($\Delta C_\ell=
\Delta C_\ell^{(3)}$, blue curve) for the linear power spectrum (dashed parts are negative values). The straight black line is the detectability threshold $\Delta C_\ell/\sigma_\ell=1$. The post-Born corrections are below the cosmic variance by two orders of magnitude up to $\ell=2500$.}
\label{Cosmic_Variance}
\end{figure}

Finally, in Fig.~\ref{Comparing-all} we plot the corrections to the unlensed $C_\ell^\Mcal$'s due to resummed $\th{1}{a}$, post-Born contribution from Eqs.~(\ref{eq:secondgroup1}) and~(\ref{eq:secondgroup2}), and post-Born non-Gaussian contribution from 
Eqs. (\ref{eq:thirdgroup1}) and (\ref{eq:thirdgroup2}), 
for both linear power spectrum and Halofit model.
As one can see the non-Gaussian part (third group) is always the most relevant post-Born correction
 in going beyond the first order deflection angles contribution. Nevertheless, these post-Born corrections are below cosmic variance, as we can see from Fig.~\ref{Cosmic_Variance}. There we plot the ratio $\Delta C_\ell/\sigma_\ell$, where
\beq
\sigma_\ell\equiv\sqrt{\frac{2}{2\ell+1}}\,C^\Mcal_\ell
\eeq
is the  cosmic variance. We have taken into account only the linear power spectrum because we are just interested in comparing the orders of magnitude of these different effects. Lensing corrections are detectable if this ratio is larger than 1. Within the range in $\ell$ where our approximations hold ($\ell\leq2500$),  post-Born corrections are smaller than  cosmic variance  by two orders of magnitude. In order to pass this threshold one would have to combine several hundred $\ell$-values into one bin.
 

\section{Conclusions}
\label{Sec7}
\setcounter{equation}{0}
In this work we have evaluated the impact of going beyond the Born approximation on the lensed CMB temperature 
 power spectrum. We postpone the study  of this effect on E and B-mode polarization for a forthcoming paper.

The evaluation of the post-Born lensing correction on the unlensed $C_\ell^{\Mcal}$ is  performed
in the small deflection angle approximation. This has the drawback that it is  reliable only for multipole $\ell\lesssim 2500$, but it allows us to consistently take into account the non-Gaussian nature of cosmological perturbation theory beyond the linear level. 
In fact, in the previous literature \cite{Hagstotz:2014qea}, the evaluation of post-Born correction was performed 
using a non-perturbative re-summation based on the hypothesis that also the non-linear part of the deflection angle $\delta \boldsymbol{\theta}$ has Gaussian statistics, needed to use  the relation 
$\langle e^{i y} \rangle=e^{-\langle y^2 \rangle/2}$ valid only for a Gaussian stochastic variable  $y$.
This is of course no longer valid when we go beyond linear order, to take into account post-Born corrections to the deflection angle.

The contributions to the lensed temperature power spectrum coming from the non-Gaussian nature of the deflection angle are given in Eqs.~(\ref{eq:thirdgroup1}) and~(\ref{eq:thirdgroup2}). This is a  new effect not taken into account in the past literature, and it turns out to be 
the leading contribution when compared with the contribution given in Eqs.~(\ref{eq:secondgroup1}) and~(\ref{eq:secondgroup2}), i.e., the contribution that will be appear also in the re-summation performed in~\cite{Hagstotz:2014qea}.

From a quantitative point of view the non-Gaussian contribution from Eqs.~(\ref{eq:thirdgroup1}) and~(\ref{eq:thirdgroup2}) is more than one 
order of magnitude larger of the one given in Eqs.~(\ref{eq:secondgroup1}) and~(\ref{eq:secondgroup2}), see Fig.~\ref{Comparing-all}.
This gives a correction that oscillates and becomes close to $0.05\%$ for multipoles  between 2000 and 2500.
This result corrects the one of~\cite{Hagstotz:2014qea}, which is on the one hand incomplete and on the other hand overestimates the contributions from the second group by about two order of magnitude.
(The fact that the result by~\cite{Hagstotz:2014qea} is similar to the one from Eqs.~(\ref{eq:secondgroup1}) and (\ref{eq:secondgroup2}) without cancellation, hints to the fact that  the overestimation is probably due to the sign error pointed out in Sect.~\ref{Sec2}.) 

Looking at Fig.~\ref{All-effects} we also note that the non-linear matter power spectrum changes the lensed temperature spectrum much more (by about $1\%$ for $\ell\sim 2500$), than the post-Born corrections.

The magnitude of the correction from going beyond the Born approximation evaluated here is
relatively small and, at the present, not detectable in CMB experiments.  On the other hand, its dominant contribution  originates from a new 
effect, neglected in~\cite{Hagstotz:2014qea}, and similar contributions could be more significant, in particular, for  B-mode 
polarization. This is  the subject of a future paper.
\vspace{0.5cm}

{\bf Note Added.}
While we were finalizing this manuscript another study of the effect of post-Born lensing corrections 
on  CMB temperature anisotropies has appeared in~\cite{Pratten:2016dsm}.
Our analytical results for higher order deflection angles and the amplification matrix~\cite{Fanizza:2015swa} are in agreement with the ones used in~\cite{Pratten:2016dsm}. This means that we   also agree on
the post-Born corrections to the lensing and curl potential power spectra.
As a consequence, we also obtain results consistent with the ones presented in~\cite{Pratten:2016dsm}
considering the contribution to add to the unlensed $C_\ell^\Mcal$ coming from the terms which we call the second group, $\langle \th{1}{a}\th{3}{b} \rangle
\langle \nabla_a\Mcal \nabla_b\bar\Mcal \rangle$ and $ \langle \th{2}{a}\th{2}{b} \rangle
\langle \nabla_a\Mcal \nabla_b\bar\Mcal \rangle$,
which is the first contribution of the re-summed series probably used in~\cite{Pratten:2016dsm} to evaluate post-Born 
contribution. Regarding this contribution, we also agree with~\cite{Pratten:2016dsm} that
some mistakes are present in~\cite{Hagstotz:2014qea} which overestimates the correction.

On the other hand, like~\cite{Hagstotz:2014qea}, also in~\cite{Pratten:2016dsm} the assumption that the lensing potential is Gaussian leads the authors to neglect contributions coming 
from an odd number of deflection angles.
In fact, despite mentioning that non-Gaussianity from post-Born corrections can give an additional contribution, this contribution is not evaluated in \cite{Pratten:2016dsm}.
Due to the non-Gaussian nature of higher order perturbations this contribution does not vanish beyond linear order and is given by the terms which we call third group, $\langle \th{1}{a}\th{1}{b}\th{2}{c} \rangle \langle \nabla_a\Mcal \nabla_b\nabla_c\bar\Mcal \rangle$ and $ \langle \th{2}{a}\th{1}{b}\th{1}{c} \rangle
 \langle \nabla_a\Mcal \nabla_b\nabla_c\bar\Mcal \rangle$.
 This leads in~\cite{Pratten:2016dsm} to an underestimation 
of the total correction from post-Born effects by at least one order of magnitude.

\section*{Acknowledgements}

We wish to thank Francesco Montanari for several useful discussions and suggestions.
We are grateful to the Swiss National Science Foundation for financial support.
GM wishes to thank CNPq for financial support.
GM was also partially supported by the Marie Curie IEF, Project NeBRiC - ``Non-linear effects and backreaction in classical and quantum cosmology" at the initial stage of this work.
GF is supported in part by MIUR, under grant no. 2012CPPYP7 
(PRIN 2012), and by INFN,  under the program TAsP ({Theoretical Astroparticle Physics}). 
GF was also supported by the foundation ``Angelo Della Riccia''.
ED is supported by the ERC Starting Grant cosmoIGM and by INFN/PD51 INDARK grant.
We thank the Galileo Galilei Institute for Theoretical Physics for the hospitality and the INFN for partial support during the completion of this work.


\appendix

\section{$\Acal^{(i....)}(\vl{})$ and $C_\ell^{(i....,j....)}$ terms}
\label{a:Al}

In $\vl{}-$space we obtain the following useful expressions for the lensing corrections to the CMB temperature fluctuations up to forth order:
\bea
\Acal^{(1)}(\vl{})&=&\frac{1}{2\pi}\int d^2 x\,\T{a}{1}\nabla_a\Mcal\,e^{i\vl{}\cdot{\bf x}}\nonumber\\
&=&\frac{1}{\pi}\INT{2}\left[\left(\vl{}-\vl{2}\right)\cdot\vl{2}\right]
\int_{0}^{r_s}dr' \frac{r_s-r'}{r_s\,r'}\,
\Phi_W(r',\vl{}-\vl{2})\Mcal(r_s,\vl{2})\,,
\\
& & \nonumber \\
&& \nonumber \\
\Acal^{(2)}(\vl{})&=&\frac{1}{2\pi}\int d^2 x\,\T{a}{2}\nabla_a\Mcal\,e^{i\vl{}\cdot{\bf x}}\nonumber\\
&=&-\frac{1}{\pi^2}\INT{2}\INT{3}
\,\left[\left(\vl{}+\vl{2}-\vl{3}\right)\cdot\vl{3}\right]\,
\left[\left(\vl{}+\vl{2}-\vl{3}\right)\cdot\vl{2}\right]\int_{0}^{r_s}dr'\frac{r_s-r'}{r_s\,r'}\nonumber\\
&&\times\int_{0}^{r'}dr'' \frac{r'-r''}{r'\,r''}\,
\Phi_W(r',\vl{}+\vl{2}-\vl{3})
\bar\Phi_W(r'',\vl{2})\Mcal(r_s,\vl{3})\,,
\\
& & \nonumber \\
&& \nonumber \\
\Acal^{(11)}(\vl{})&=&\frac{1}{2\pi}\int d^2 x\frac{1}{2}\T{a}{1}\T{b}{1}\nabla_a\nabla_b\Mcal e^{i\vl{}\cdot {\bf x}}\nonumber\\
&=&-\frac{1}{2}\frac{1}{\pi^2}\INT{2}\INT{3}\,\left[\left(\vl{}+\vl{2}-\vl{3}\right)\cdot \vl{3}\right]\,\left(\vl{2}\cdot \vl{3}\right)\nonumber\\
&&\times\int_{0}^{r_s}dr' \frac{r_s-r'}{r_s\,r'}\int_{0}^{r_s}dr'' \frac{r_s-r''}{r_s\,r''}
\Phi_W(r',\vl{}+\vl{2}-\vl{3})
\bar\Phi_W(r'',\vl{2})\,\Mcal(r_s,\vl{3})\,,\nonumber \\
\\
& & \nonumber \\
\Acal^{(3)}(\vl{})
&=&\frac{1}{2\pi}\int d^2 x\,\T{a}{3}\nabla_a\Mcal\,e^{i\vl{}\cdot{\bf x}}\nonumber\\
&=&
\frac{1}{\pi^3}\INT{2}\INT{3}\INT{4}\left\{ \left[\left(\vl{}-\vl{2}-\vl{3}-\vl{4}\right)\cdot\vl{4}\right]\,\left[\left(\vl{}-\vl{2}-\vl{3}-\vl{4}\right)\cdot\vl{2}\right]\right.\nonumber \\
&&\left. \times \left(\vl{2}\cdot\vl{3}\right)
\int_{0}^{r_s}dr'\frac{r_s-r'}{r_s\,r'}\int_{0}^{r'}dr''\frac{r'-r''}{r'\,r''}
\int_{0}^{r''}dr''' \frac{r''-r'''}{r''\,r'''}\right.\nonumber\\
&& \times\,\Phi_W(r',\vl{}-\vl{2}-\vl{3}-\vl{4})\Phi_W(r'',\vl{2})\Phi_W(r''',\vl{3})\Mcal(r_s,\vl{4})\nonumber\\
&&\left.+\frac{1}{2}\left[\left(\vl{}-\vl{2}-\vl{3}-\vl{4}\right)\cdot\vl{4}\right]\,\left[\left(\vl{}-\vl{2}-\vl{3}-\vl{4}\right)\cdot\vl{2}\right]\,
\left[\left(\vl{}-\vl{2}-\vl{3}-\vl{4}\right)\cdot\vl{3}\right]\right.\nonumber\\
&&\left.\times\int_{0}^{r_s}dr'\frac{r_s-r'}{r_s\,r'}
\int_{0}^{r'}dr'' \frac{r'-r''}{r'\,r''}\,
\int_{0}^{r'}dr''' \frac{r'-r'''}{r'\,r'''}\right.\nonumber\\
&& \left.\times\,\Phi_W(r',\vl{}-\vl{2}-\vl{3}-\vl{4})\Phi_W(r'',\vl{2})\Phi_W(r''',\vl{3})\Mcal(r_s,\vl{4})\right\}\,,
\eea
\bea
\Acal^{(12)}(\vl{})
&=&\frac{1}{2\pi}\int d^2 x\,\T{a}{1}\T{b}{2}\nabla_a\nabla_b\Mcal\,e^{i\vl\cdot{\bf x}}\nonumber\\
&=&\frac{1}{\pi^3}
\INT{2}\INT{3}\INT{4}\left[\left(\vl{}-\vl{2}-\vl{3}-\vl{4}\right)\cdot\vl{4}\right]\,
\left(\vl{4}\cdot\vl{2}\right)\,
\left(\vl{3}\cdot\vl{2}\right)\nonumber\\
&&\times\int_{0}^{r_s}dr' \frac{r_s-r'}{r_s\,r'}\int_{0}^{r_s}dr''\frac{r_s-r''}{r_s\,r''}
\int_{0}^{r''}dr''' \frac{r''-r'''}{r''\,r'''}\,\nonumber
\\
&&\times\,\Phi_W(r',\vl{}-\vl{2}-\vl{3}-\vl{4})\Phi_W(r'',\vl{2})\Phi_W(r''',\vl{3})\Mcal(r_s,\vl{4})\,,
\\
& & \nonumber \\
&& \nonumber \\
\Acal^{(111)}(\vl{})
&=&\frac{1}{2\pi}\int d^2 x\frac{1}{6}\,\T{a}{1}\T{b}{1}\T{c}{1}\nabla_a\nabla_b\nabla_c\Mcal\,e^{i\vl{}\cdot{\bf x}}\nonumber\\
&=&
\frac{1}{6}\frac{1}{\pi^3}\INT{2}\INT{3}\INT{4}\left[\left(\vl{}-\vl{2}-\vl{3}-\vl{4}\right)\cdot\vl{4}\right]\,
\left(\vl{2}\cdot\vl{4}\right)\,
\left(\vl{3}\cdot\vl{4}\right)
\nonumber\\
&&\times
\int_{0}^{r_s}dr' \frac{r_s-r'}{r_s\,r'}\,\int_{0}^{r_s}dr'' \frac{r_s-r''}{r_s\,r''}\,
\int_{0}^{r_s}dr''' \frac{r_s-r'''}{r_s\,r'''}\nonumber\\
&&\times\,\Phi_W(r',\vl{}-\vl{2}-\vl{3}-\vl{4})\Phi_W(r'',\vl{2})\Phi_W(r''',\vl{3})\Mcal(z_s,\vl{4})\,,
\\
&& \nonumber \\
&& \nonumber \\
\Acal^{(22)}(\vl{})
&=&\frac{1}{2\pi}\int d^2 x\,\frac{1}{2}\,\T{a}{2}\T{b}{2}\nabla_a\nabla_b\Mcal\,e^{i \vl\cdot{\bf x}}\nonumber\\
&=&\frac{1}{2}\frac{1}{\pi^4}\INT{2}\INT{3}\INT{4}\INT{5}
\left[\left(\vl{}-\vl{2}-\vl{3}-\vl{4}-\vl{5}\right)\cdot\vl{5}\right]\nonumber\\
&&\times\left[\left(\vl{}-\vl{2}-\vl{3}-\vl{4}-\vl{5}\right)\cdot\vl{2}\right]
\left(\vl{5}\cdot\vl{3}\right)\,
\left(\vl{3}\cdot\vl{4}\right)
\int_{0}^{r_s}dr'\frac{r_s-r'}{r_s\,r'}\nonumber\\
&&\times\int_{0}^{r'}dr'' \frac{r'-r''}{r'\,r''}\,
\int_{0}^{r_s}dr'''\frac{r_s-r'''}{r_s\,r'''}
\int_{0}^{r'''}dr'''' \frac{r'''-r''''}{r'''\,r''''}\nonumber\\
&&\times\,\Phi_W(r',\vl{}-\vl{2}-\vl{3}-\vl{4}-\vl{5})\Phi_W(r'',\vl{2})\Phi_W(r''',\vl{3})\Phi_W(r'''',\vl{4})\Mcal(r_s,\vl{5})\,,
\nonumber \\
\\
&& \nonumber \\
\Acal^{(13)}(\vl{})
&=&\frac{1}{2\pi}\int d^2 x\,\T{a}{1}\T{b}{3}\nabla_a\nabla_b\Mcal\,e^{i\vl{}\cdot{\bf x}}\nonumber\\
&=&\frac{1}{\pi^4}
\INT{2}\INT{3}\INT{4}\INT{5}
\left\{\left[\left(\vl{}-\vl{2}-\vl{3}-\vl{4}-\vl{5}\right)\cdot\vl{5}\right]\,
\left(\vl{2}\cdot\vl{5}\right)\right.\nonumber\\
&&\times
\left(\vl{2}\cdot\vl{3}\right)
\left(\vl{3}\cdot\vl{4}\right) 
\int_{0}^{r_s}dr' \frac{r_s-r'}{r_s\,r'}\int_{0}^{r_s}dr''\frac{r_s-r''}{r_s\,r''}
\int_{0}^{r''}dr'''\frac{r''-r'''}{r''\,r'''}
\nonumber\\
&&\times\int_{0}^{r'''}dr'''' \frac{r'''-r''''}{r'''\,r''''}\Phi_W(r',\vl{}-\vl{2}-\vl{3}-\vl{4}-\vl{5})\Phi_W(r'',\vl{2})\nonumber\\
& & \times \Phi_W(r''',\vl{3})\Phi_W(r'''',\vl{4})\Mcal(r_s,\vl{5})\nonumber \\
&&+\frac{1}{2}\left[\left(\vl{}-\vl{2}-\vl{3}-\vl{4}-\vl{5}\right)\cdot\vl{5}\right]\,
\left(\vl{2}\cdot\vl{5}\right)
\left(\vl{2}\cdot\vl{3}\right)\,
\left(\vl{2}\cdot\vl{4}\right)\int_{0}^{r_s}dr' \frac{r_s-r'}{r_s\,r'}\,
\nonumber\\
&&\times\int_{0}^{r_s}dr''\frac{r_s-r''}{r_s\,r''}
\int_{0}^{r''}dr''' \frac{r''-r'''}{r''\,r'''}\,
\int_{0}^{r''}dr'''' \frac{r''-r''''}{r''\,r''''}\nonumber\\
&&\left.\times\,
\Phi_W(r',\vl{}-\vl{2}-\vl{3}-\vl{4}-\vl{5})\Phi_W(r'',\vl{2})\Phi_W(r''',\vl{3})\Phi_W(r'''',\vl{4})\Mcal(r_s,\vl{5})\right\}\,,
\nonumber\\
&&
\eea
\bea
\Acal^{(112)}(\vl{})
&=&\frac{1}{2\pi}\int d^2 x\,\frac{1}{2}\T{a}{1}\T{b}{1}\T{c}{2}\nabla_a\nabla_b\nabla_c\Mcal\,e^{i\vl{}\cdot{\bf x}}\nonumber\\
&=&-\frac{1}{2}\frac{1}{\pi^4}
\INT{2}\INT{3}\INT{4}\INT{5}
\left[\left(\vl{}-\vl{2}-\vl{3}-\vl{4}-\vl{5}\right)\cdot\vl{5}\right]\,
\left(\vl{2}\cdot\vl{5}\right)\nonumber\\
&&\times\left(\vl{5}\cdot\vl{3}\right)\,
\left(\vl{3}\cdot\vl{4}\right)
\int_{0}^{r_s}dr' \frac{r_s-r'}{r_s\,r'}
\int_{0}^{r_s}dr'' \frac{r_s-r''}{r_s\,r''}\nonumber\\
&&\times
\int_{0}^{r_s}dr'''\frac{r_s-r'''}{r_s\,r'''}
\int_{0}^{r'''}dr'''' \frac{r'-r''''}{r'\,r''''}\nonumber
\\
&&\times\Phi_W(r',\vl{}-\vl{2}-\vl{3}-\vl{4}-\vl{5})\Phi_W(r'',\vl{2})\Phi_W(r''',\vl{3})\Phi_W(r'''',\vl{4})\Mcal(r_s,\vl{5})\,,
\nonumber
\\
&& \\
&&\nonumber \\
\Acal^{(1111)}(\vl{})
&=&\frac{1}{2\pi}\int d^2 x\,\frac{1}{24}\,\T{a}{1}\T{b}{1}\T{c}{1}\T{d}{1}\nabla_a\nabla_b\nabla_c\nabla_d\Mcal\,e^{i\vl{}\cdot{\bf x}}\nonumber\\
&=&
\frac{1}{24}\frac{1}{\pi^4}
\INT{2}\INT{3}\INT{4}\INT{5}\left[\left(\vl{}-\vl{2}-\vl{3}-\vl{4}-\vl{5}\right)\cdot\vl{5}\right]
\nonumber\\
&&
\times 
\left(\vl{2}\cdot\vl{5}\right)\,
\left(\vl{3}\cdot\vl{5}\right)\,
\left(\vl{4}\cdot\vl{5}\right)
\int_{0}^{r_s}dr' \frac{r_s-r'}{r_s\,r'}\int_{0}^{r_s}dr'' \frac{r_s-r''}{r_s\,r''}\,
\int_{0}^{r_s}dr''' \frac{r_s-r'''}{r_s\,r'''}\nonumber\\
&&
\times
\int_{0}^{r_s}dr'''' \frac{r_s-r''''}{r_s\,r''''}
\Phi_W(r',\vl{}-\vl{2}-\vl{3}-\vl{4}-\vl{5})\Phi_W(r'',\vl{2})\Phi_W(r''',\vl{3})\nonumber \\
&&
\times\Phi_W(r'''',\vl{4})\Mcal(r_s,\vl{5})\,.
\eea
We do not write  the term associated to $\theta^{a(4)}_s$ because its contribution to the angular power spectrum of lensed CMB temperature anisotropies vanishes at fourth order. This is a consequence of statistical isotropy  analogous to what happens to the second order contribution coming from $\theta^{a(2)}_s$, which has been shown to vanish in~\cite{Bonvin:2015uha}.

The above expressions can be used to calculate the higher order lensing corrections to the temperature power spectrum (see Eq. \ref{HigherOrderCll}). After some algebra, and omitting contributions that vanish after angular integrations, we 
obtain 

\bea
C_\ell^{(0,13)} &=&
-8\,C_{\ell}^\Mcal(z_s)
\NINT{1}\NINT{2}\left(\vl{1}\cdot\vl{}\right)\,
\left(\vl{2}\cdot\vl{}\right)\,
\ell_2^2\,
\left(\vl{2}\cdot\vl{1}\right)\,\nonumber\\
&&\times
\int_{0}^{r_s}dr'\frac{r_s-r'}{r_s\,r'}
\int_{0}^{r'}dr''\frac{r'-r''}{r'\,r''}\,
C_{\ell_1}^\psi(z_s,z'')C_{\ell_2}^W(z'',z')
\nonumber 
\\
& &
+8\,C_{\ell}^\Mcal(z_s)
\NINT{1}\NINT{2}
\left(\vl{2}\cdot\vl{}\right)\,
\ell_2^2\,
\left(\vl{2}\cdot\vl{1}\right)
\left(\vl{1}\cdot\vl{}\right)
\nonumber\\
&&\times
\int_{0}^{r_s}dr'\frac{r_s-r'}{r_s\,r'}
C_{\ell_1}^\psi(z_s,z')\,
\int_{0}^{r'}dr'' \frac{r'-r''}{r'\,r''}
C_{\ell_2}^W(z'',z')
\nonumber 
\\
& &
+4\,C_{\ell}^\Mcal(z_s)
\NINT{1}\NINT{2}
\left(\vl{1}\cdot\vl{}\right)^2\,
\left(\vl{1}\cdot\vl{2}\right)^2
\nonumber
\\
&&\times
\int_{0}^{r_s}dr' \frac{r_s-r'}{r_s\,r'}\,
\int_{0}^{r_s}dr''\frac{r_s-r''}{r_s\,r''}
C_{\ell_1}^W(z',z'')C_{\ell_2}^\psi(z'',z'') \,,   
\label{C13}
\eea
\bea
C_\ell^{(0,22)} &=&
-4\,C_{\ell}^\Mcal(z_s)
\NINT{1}\NINT{2}\left(\vl{}\cdot\vl{1}\right)^2\,
\left(\vl{1}\cdot\vl{2}\right)^2\,
\nonumber
\\
&&\times
\int_{0}^{r_s}dr'\frac{r_s-r'}{r_s\,r'}
\int_{0}^{r_s}dr''\frac{r_s-r''}{r_s\,r''}
C_{\ell_1}^W(z',z'')C_{\ell_2}^\psi(z',z'')\,, 
\label{C22}
\\
& & \nonumber \\
& & \nonumber \\
C_\ell^{(0,1111)} &=&
\frac{1}{4}C_{\ell}^\Mcal(z_s)\left[
\NINT{1}\,\left(\vl{1}\cdot\vl{}\right)^2
C^\psi_{\ell_1}(z_s,z_s)\right]^2\,, 
\\
& & \nonumber \\
& & \nonumber \\
C_\ell^{(1, 3)} &=&
8\,\NINT{1}\NINT{2}\left(\vl{2}\cdot\vl{1}\right)\ell_2^2\,
\left[\left(\vl{}-\vl{1}\right)\cdot\vl{2}\right]
\left[\left(\vl{}-\vl{1}\right)\cdot\vl{1}\right]\,
C_{\ell_1}^\Mcal(z_s)\,
\nonumber
\\
&&
\times
\int_{0}^{r_s}dr'\frac{r_s-r'}{r_s\,r'}
\int_{0}^{r'}dr''\frac{r'-r''}{r'\,r''} 
C_{\ell_2}^W(z',z'')C_{|\vl{}-\vl{1}|}^\psi(z'',z_s)
\nonumber
\\
& &
-8\,\NINT{1}\NINT{2}\left[\left(\vl{}-\vl{1}\right)\cdot\vl{1}\right]\,
\left(\vl{2}\cdot\vl{1}\right)\ell_2^2\,
\left[\left(\vl{}-\vl{1}\right)\cdot\vl{2}\right]\,
C_{\ell_1}^\Mcal(z_s)\,
\nonumber \\
&&
\times
\int_{0}^{r_s}dr'\frac{r_s-r'}{r_s\,r'}
C_{|\vl{}-\vl{1}|}^\psi(z',z_s)
\int_{0}^{r'}dr'' \frac{r'-r''}{r'\,r''}
C_{\ell_2}^W(z',z'')
\nonumber \\
& &
-4\,\NINT{1}\NINT{2}
\left[\left(\vl{}-\vl{1}\right)\cdot\vl{1}\right]^2\,
\left[\left(\vl{}-\vl{1}\right)\cdot\vl{2}\right]^2\,
C_{\ell_1}^\Mcal(z_s)\,
\nonumber\\
&&
\times
\int_{0}^{r_s}dr'\frac{r_s-r'}{r_s\,r'}
C_{\ell_2}^\psi(z',z')
\int_{0}^{r_s}d\eta'' \frac{r_s-r''}{r_s\,r''}
C_{|\vl{}-\vl{1}|}^W(z',z'') \,,    
\\
& & \nonumber \\
& & \nonumber \\
C_\ell^{(2, 2)} &=&
4\,\NINT{1}\NINT{2}\,\left[\left(\vl{}+\vl{2}-\vl{1}\right)\cdot\vl{1}\right]^2\,
\left[\left(\vl{}+\vl{2}-\vl{1}\right)\cdot\vl{2}\right]^2\,C_{\ell_1}^\Mcal(z_s)
\nonumber\\
&&
\times
\int_{0}^{r_s}dr'\frac{r_s-r'}{r_s\,r'}\,\int_{0}^{r_s}dr''\frac{r_s-r''}{r_s\,r''}\,
C_{|\vl{}+\vl{2}-\vl{1}|}^W(z',z'')C_{\ell_2}^\psi(z'',z')
\nonumber
\\
& &
-16\,
\NINT{1}\NINT{2}\,\left(\vl{1}\cdot\vl{2}\right)\,
\left[\left(\vl{}+\vl{1}-\vl{2}\right)\cdot\vl{2}\right]\,
\left[\left(\vl{}+\vl{1}-\vl{2}\right)\cdot\vl{1}\right]^2\nonumber\\
&&\times
C_{\ell_2}^\Mcal(z_s)
\int_{0}^{r_s}dr'\frac{r_s-r'}{r_s\,r'}
\int_{0}^{r'}dr'' \frac{r'-r''}{r'\,r''}\nonumber\\
&&\times
\int_{0}^{r_s}dr'''\frac{r_s-r'''}{r_s\,'''}\,
C_{\ell_1}^W(z'',z''') 
\int_{0}^{r'''}dr'''' \frac{r'''-r''''}{r'''\,r''''}
C_{|\vl{}+\vl{1}-\vl{2}|}^W(z',z'''') \,, \nonumber \\
& & 
\\
& & \nonumber \\
C_\ell^{(1, 12)} &=&
-8
\NINT{1}\NINT{2}
\left(\vl{2}\cdot\vl{1}\right)\,
\left[(\vl{}-\vl{1})\cdot\vl{2}\right]\,
\left[\left(\vl{}-\vl{1}\right)\cdot\vl{1}\right]^2C_{\ell_1}^\Mcal(z_s)
\nonumber\\
&&
\times
\int_{0}^{r_s}dr'\frac{r_s-r'}{r_s\,r'}C_{\ell_2}^\psi(z_s,z')\,
\int_{0}^{r_s}dr'' \frac{r_s-r''}{r_s\,r''}
C_{|\vl{}-\vl{1}|}^W(z'',z') \,,   
\eea
\bea
C_\ell^{(1, 111)} &=&
-\NINT{1}\NINT{2}\left[\left(\vl{}-\vl{1}\right)\cdot\vl{1}\right]^2\left(\vl{2}\cdot\vl{1}\right)^2\,
C_{\ell_1}^\Mcal(z_s)
C_{|\vl{}-\vl{1}|}^\psi(z_s,z_s)C_{\ell_2}^\psi(z_s,z_s) \,, \nonumber\\
&&
\\
& & \nonumber \\
C_\ell^{(2, 11)} &=&
8\, \NINT{1}\NINT{2}\,
\left[\left(\vl{}+\vl{1}-\vl{2}\right)\cdot \vl{2}\right]^2\,
\left(\vl{1}\cdot \vl{2}\right)\,
\left[\left(\vl{}+\vl{1}-\vl{2}\right)\cdot\vl{1}\right] \nonumber\\
&&
\times
C_{\ell_2}^\Mcal(z_s)\int_{0}^{r_s}dr'\frac{r_s-r'}{r_s\,r'}
C_{\ell_1}^\psi(z_s,z')\int_{0}^{r_s}dr'' \frac{r_s-r''}{r_s\,r''}C_{|\vl{}+\vl{1}-\vl{2}|}^W(z',z'')  \,, 
\\
& & \nonumber \\
& & \nonumber \\
C_\ell^{(11, 11)} &=&
C_{\ell}^\Mcal(z_s)
\left[\frac{1}{2}\frac{1}{(2\pi)^2}
\INT{1}\,\left(\vl{1}\cdot \vl{}\right)^2\,C_{\ell_1}^\psi(z_s,z_s)\right]^2
\nonumber \\
& &
+\frac{1}{2}\NINT{1}\NINT{2}\,\left\{\left[\left(\vl{}+\vl{1}-\vl{2}\right)\cdot \vl{2}\right]\,
\left(\vl{1}\cdot \vl{2}\right)\right\}^2\,
C_{\ell_2}^\Mcal(z_s) \hspace{3.5cm} \nonumber\\
&&
\times
C_{|\vl{}+\vl{1}-\vl{2}|}^\psi(z_s,z_s)C_{\ell_
1}^\psi(z_s,z_s) \,.
\eea


\bibliographystyle{JHEP}
\bibliography{biblio_CMBlensing}

\end{document}